\documentclass[aps, prd, twocolumn,superscriptaddress,nofootinbib,10pt]{revtex4-2}
\usepackage{graphicx}
\usepackage{dcolumn}
\usepackage{multirow}
\usepackage[table]{xcolor}
\usepackage{colortbl,hhline}
\usepackage{amssymb,amsmath,amsthm,mathrsfs}
\usepackage{epstopdf}
\usepackage{hyperref}
\usepackage{empheq}
\usepackage{color}
\usepackage[normalem]{ulem} 
\usepackage{physics}
\usepackage[normalem]{ulem}
\usepackage[ruled,vlined]{algorithm2e}
\usepackage{enumitem}
\usepackage{empheq}
\usepackage{comment}
\usepackage{hyperref}
\usepackage{cleveref}
\usepackage{placeins}

\usepackage[outercaption]{sidecap}
\newcommand{\be}{\begin{equation}}
\newcommand{\ee}{\end{equation}}
 \newcommand{\bea}{\begin{eqnarray}}
\newcommand{\eea}{\end{eqnarray}}

\usepackage{tikz}
\usetikzlibrary{calc,3d,arrows}
\usetikzlibrary{decorations.pathreplacing}
\usetikzlibrary{arrows.meta}

\newcommand{\mpl}{m_{{}_{\mathrm{p}}}}

\definecolor{burntorange}{rgb}{0.8, 0.33, 0.0}

\begin{document}

\title{Gravitational wave and particle emission from a cosmic string loop: local case}

\newcommand{\addressIFIC}{Instituto de F\'isica Corpuscular (IFIC), Consejo Superior de Investigaciones Cient\'ificas (CSIC) and Universitat de Val\`{e}ncia, 46980, Valencia, Spain}
\newcommand{\addressEHU}{Department of Physics, University of Basque Country, UPV/EHU, 48080, Bilbao, Spain}
\newcommand{\addressEHUQC}{EHU Quantum Center, University of the Basque Country UPV/EHU, Leioa, 48940 Biscay, Spain}
\newcommand{\addressNottingham}{School of Physics and Astronomy, University of Nottingham, Nottingham, NG7 2RD, UK}

\author{Jorge Baeza-Ballesteros}\email{jorge.baeza@ific.uv.es} \affiliation{\addressIFIC} 
\author{Edmund J. Copeland}\email{edmund.copeland@nottingham.ac.uk} \affiliation{\addressNottingham}

\author{Daniel G. Figueroa} \email{daniel.figueroa@ific.uv.es} \affiliation{\addressIFIC} 
\author{Joanes Lizarraga}\email{joanes.lizarraga@ehu.eus} \affiliation{\addressEHU}\affiliation{\addressEHUQC}

\date{\today}

\begin{abstract}
Using lattice field simulations of the Abelian-Higgs model, we characterize the simultaneous emission of (scalar and gauge) particles and gravitational waves (GWs) by local string loops. We use {\it network} loops created in a phase transition, and {\it artificial} loops formed by either crossing straight-boosted or curved-static infinite strings. Loops decay via both particle and GW emission, on time scales $\Delta t_{\rm dec} \propto L^p$, where $L$ is the loop length. For particle production, we find $p \simeq 2$ for artificial loops and $p \simeq 1$ for network loops, whilst for GW emission, we find $p \simeq 1$ for all loops. We find that below a critical length, artificial loops decay primarily through particle production, whilst for larger loops GW emission dominates. However, for network loops, which represent more realistic configurations, particle emission always dominates, as supported by our data with length-to-core ratios up to $L/r_\text{c} \lesssim 6000$. Our results indicate that the GW background from a local string network should be greatly suppressed compared to estimations that ignore particle emission.
\end{abstract}


\maketitle

\section{Introduction}

It is nearly half a century since cosmic strings were introduced by Tom Kibble~\cite{Kibble:1976sj}. These are 
topologically stable line-like configurations predicted to form in the early universe by many extension of the Standard Model~\cite{Kibble:1976sj,Kibble:1980mv,Vilenkin:1984ib,Hindmarsh:1994re,Copeland:2009ga,Copeland:2011dx,Vachaspati:2015cma}, forming a network of strings composed of `long’ or infinite strings, stretching across the 
horizon, and loops. The long-string density decreases
as they intercommute forming loops, which in turn eventually decay. However, the dominant decay route of the loops
has been a matter of debate 
since Kibble’s  pioneering paper. 

In the Nambu-Goto (NG) approximation of infinitely-thin strings, the loops decay solely into gravitational waves (GWs), leading to a 
GW background (GWB)~\cite{Vilenkin:1981bx,Hogan:1984is,Vachaspati:1984gt} potentially observable~\cite{Blanco-Pillado:2017rnf,Blanco-Pillado:2017oxo,Auclair:2019wcv,Gouttenoire:2019kij,Blanco-Pillado:2024aca}, depending on the string tension $\mu$. 
In reality, cosmic strings arise out of a phase transition involving a scalar-gauge sector, and hence a separate decay route opens up, as loops may decay into the fields they are made of~\cite{Hindmarsh:2017qff,Hindmarsh:2021mnl}. While this was a moot point in the past, 
given the string's negligible contribution to Cosmic Microwave Background (CMB) anisotropies~\cite{Ade:2013xla,Lazanu:2014eya,Lizarraga:2014eaa,Lizarraga:2014xza,Charnock:2016nzm,Lizarraga:2016onn}, the situation has now changed in the dawn of an era of GW cosmology. 

Pulsar timing array (PTA) collaborations have announced the first evidence for a GWB around $\sim \text{nHz}$ frequencies~\cite{NANOGrav:2023gor, Antoniadis:2023ott,Reardon:2023gzh, Xu:2023wog}. While cosmic strings are possible sources 
of such observations, their contribution depends sensitively on the type of decay strings experience: {\it is it primarily through GWs, or via particle production?} 
For example, fitting PTA data to NG cosmic strings that only emit GWs leads to the tight constraint $G\mu \lesssim 10^{-10}$~\cite{NANOGrav:2023hvm,Antoniadis:2023xlr,Figueroa:2023zhu},  with $G$ Newton’s constant, whilst fitting to a field-theory network that allows for particle production,  
loosens the constraint to 
$G \mu \lesssim 10^{-7}$~\cite{Kume:2024adn}. 

Revisiting the question of the dominant decay channel of local cosmic strings thus seems in order. With this in mind, \cite{Matsunami:2019fss} set up and evolved artificially constructed Abelian-Higgs loops, comparing the observed particle production 
with the traditional GW result from NG loops. For loops below a critical length, they found decay primarily through particle production, whilst for larger loops, GW emission dominates. This work was later complemented by \cite{Hindmarsh:2021mnl}, which found analogous results for a different kind of artificially-constructed loops, but also that particle emission is the dominant decay route for loops originating from string network of any length.

In this work, we extend the approach of~\cite{Matsunami:2019fss,Hindmarsh:2021mnl}, with one important difference: we allow for the simultaneous emission of GWs and particles, as a true comparison requires. For the comparison, we consider two-types of artificially constructed  loops  following as ~\cite{Matsunami:2019fss,Hindmarsh:2021mnl}, as well as more realistic loops from lattice-simulated networks as in~\cite{Hindmarsh:2017qff,Hindmarsh:2021mnl}. We extend in this way to local cosmic string loops our previous study for global string loops~\cite{Baeza-Ballesteros:2023say}. Our key findings, discussed below, include a result similar to \cite{Matsunami:2019fss} with regard the existence of a critical length for artificial loops. 
However, the decay of loops originating from networks is found to be always dominated by particle emission. This implies that calculations 
ignoring particle production 
overestimate the amount of GW emitted, overconstraining $G\mu$.

This work is organized as follows. In \cref{sec:model} we introduce the model used for this study, and describe the procedures used to create the different types of isolated string loops. We then present our results for the decay of loops into particles and GWs in \cref{sec:resultsdecay,sec:GWs}, respectively, and conclude in \cref{sec:conclusions}. One appendix is also included, \cref{app:simulationdetails}, where details of the  simulations  are summarized.

\section{Model and loop configuration}\label{sec:model}

We consider an Abelian-Higgs model with a complex scalar field, $\varphi=(\phi_1+i\phi_2)/\sqrt{2}$, and a U(1) gauge field, $A_\mu$, with 
Lagrangian density 
\begin{equation}
\mathcal{L} = -\left[(D_{\mu}\varphi) (D^{\mu}\varphi)^*+\frac{1}{4}F_{\mu\nu}F^{\mu\nu}+V(\varphi)\right]\,,
\end{equation} with $D_\mu=\partial_\mu-ieA_\mu$, 
$F_{\mu\nu}=\partial_\mu A_\nu-\partial_\nu A_\mu$, 
$V(\varphi)=\lambda(\varphi^*\varphi-v^2/2)^2$, $e$ the gauge coupling, and $\lambda$ and $v$ the self-coupling and vacuum expectation value ({\it vev}) of $\varphi$. After a phase transition from $\langle\varphi\rangle=0$ (symmetric phase) to $\langle|\varphi|^2\rangle=v^2/2$ (broken phase), where $\langle\cdots\rangle$ denotes the expectation value, 
cosmic strings arise 
as line-like configurations with  energy-core widths set by the inverse masses 
of the resulting 
scalar and gauge 
particles at low energies, $r_\text{s} \sim 1/m_\text{s}$ and $r_\text{g} \sim 1/m_\text{g}$, 
with $m_\text{s} = \sqrt{2\lambda}v$ and $m_\text{g} = ev$. Upper bounds on the {\it vev} are set by CMB searches~\cite{Ade:2013xla,Lazanu:2014eya,Lizarraga:2014eaa,Lizarraga:2014xza,Charnock:2016nzm,Lizarraga:2016onn} on cosmic strings as $v \lesssim 4.3\cdot10^{15}$ GeV, and from PTA searches~\cite{NANOGrav:2023hvm,Antoniadis:2023xlr,Figueroa:2023zhu}, assuming NG and $\mu=\pi v^2$, as $v \lesssim 6.9\cdot10^{13}$ GeV. In this work, we restrict ourselves to the so-called critical case, $e^2/\lambda = 2$. 

Using the  $\mathcal{C}$osmo$\mathcal{L}$attice suite~\cite{Figueroa:2020rrl, Figueroa:2021yhd,Figueroa:2023xmq} we simulate the dynamics of  the scalar-gauge field system and the GWs~\cite{GWmodule:2022,GWmodule:2023} in cubic periodic lattices. We denote as $N$, $\delta x$ and $L_\text{B}=N\delta x$ the number of sites per direction, the lattice spacing, and the physical size of the lattice, respectively. We study three loop families, which we denote as network loops and artificial loops of type I and type II. They are generated following different procedures, detailed in the subsequent subsections.

We study the time-evolution of the loop length, $L$, by 
identifying pierced plaquettes in a gauge-invariant manner~\cite{Kajantie:1998bg,Hindmarsh:2017qff}, accounting for the Manhattan effect~\cite{Vachaspati:1984dz,Rajantie:1998vv,Fleury:2015aca}. We also measure the energy of strings as
\begin{eqnarray}\label{eq:energy}
E_\text{str}=\hspace{-0.5mm}\int \hspace{-0.75mm}\text{d}^3 x \,W(\varphi) \Big[ |\dot{\varphi}|^2 +|\vec D\varphi|^2 +\frac{\vec{E}^2}{2} + \frac{\vec{B}^2}{2} + V(\varphi)\Big],\nonumber
\end{eqnarray}
where $E_i=F_{0i}$ and $B_i=\varepsilon_{ijk}F^{jk}/2$ are the electric and magnetic fields, and $W(\varphi) = [V(\varphi)/W_0]\cdot\Theta(v^2/2-|\varphi|^2)$ is a {\it weight function} that only selects regions occupied by strings, with $W_0 = \lambda v^4 / 4$ and $\Theta(x)$ the step function. 

Throughout this paper, we often express observables
in terms of dimensionless 
variables: $\tilde{\varphi}=\varphi/v$, $\tilde{L}=\sqrt{\lambda}v\,L$, $\tau = \sqrt{\lambda}v\,t$,  ${\tilde E}_{\text{str}}=(\sqrt{\lambda}/v)\,E_{\text{str}}$, and for the (comoving) momentum,  $\kappa=k/\sqrt{\lambda}v$.

\subsection{Network loops}

Network loops are generated from the decay of string networks that are close to the scaling regime~\cite{Vilenkin:1982ks,Baier:1985cn,Martins:1996jp}, following the procedure in \cite{Hindmarsh:2021mnl}. They are expected to have shapes and features similar to loops from a realistic phase transition in the early universe. Simulations are \mbox{initialized} with a Gaussian random realization of the complex scalar field in Fourier space, with power spectrum for each field component,
\begin{equation}
\Delta_{\phi_i}(k)=\frac{k^3 v^2 \ell_\text{str}^3}{\sqrt{2\pi}}\exp\left(-\frac{1}{2}k^2\ell_\text{str}^2\right)\,,
\end{equation}
normalized so that $\langle\phi_1^2+\phi_2^2\rangle=v^2$. Here $\ell_\text{str}$ is a correlation length that controls the density of the resulting network. The gauge field and the time derivatives of both fields are set to zero. 

\begin{figure*}[t!p]
    \begin{minipage}{1\textwidth}
      \centering
 		\begin{tikzpicture}[>={LaTeX[width=3mm,length=4mm]},->]
  \node[] at (-6.,0) {\includegraphics[width=0.29\textwidth]{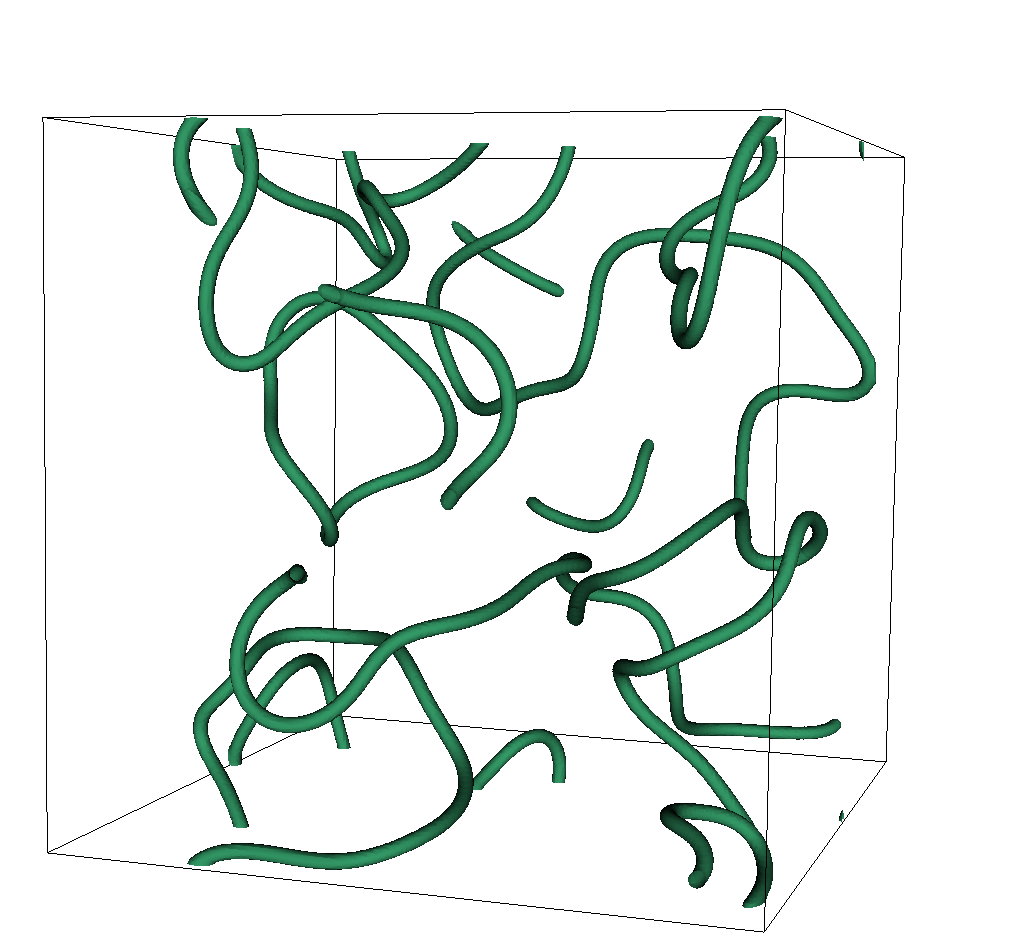}}; 
  \node[] at (0,0) {\includegraphics[width=0.29\textwidth]{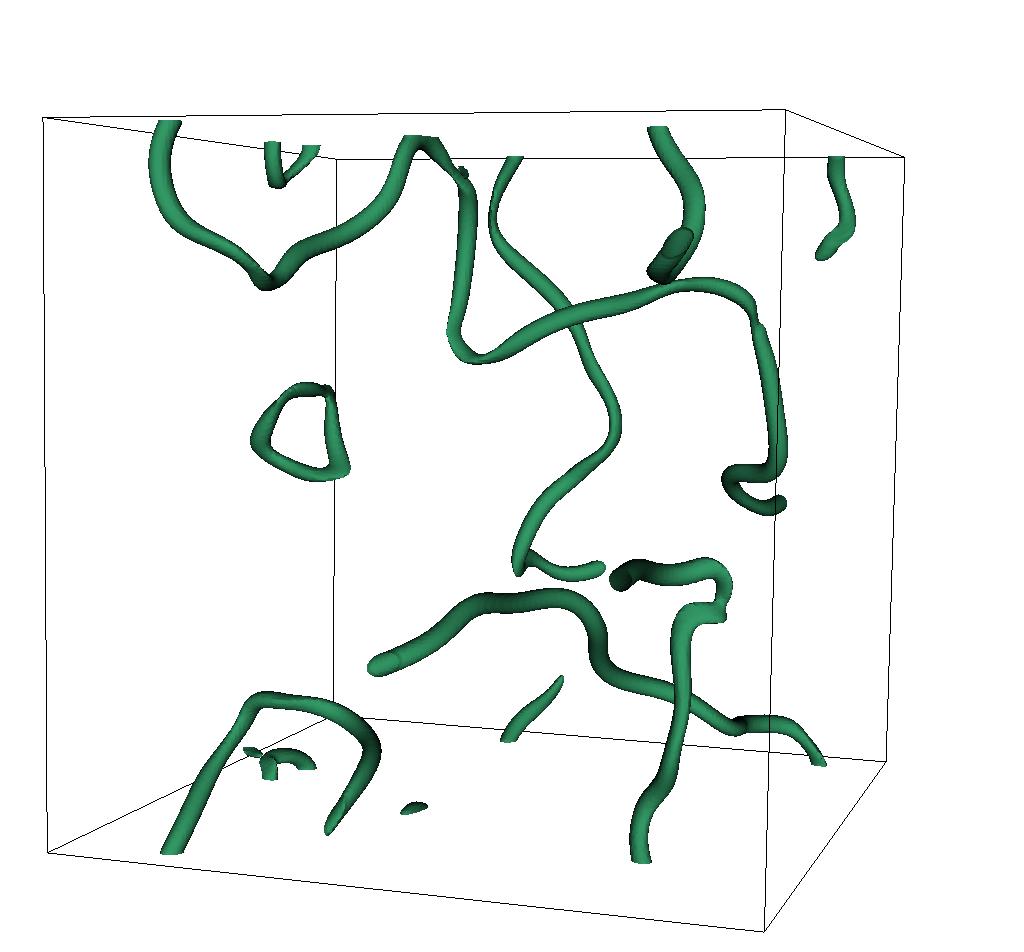}};
  \node[] at (6.,0) {\includegraphics[width=0.29\textwidth]{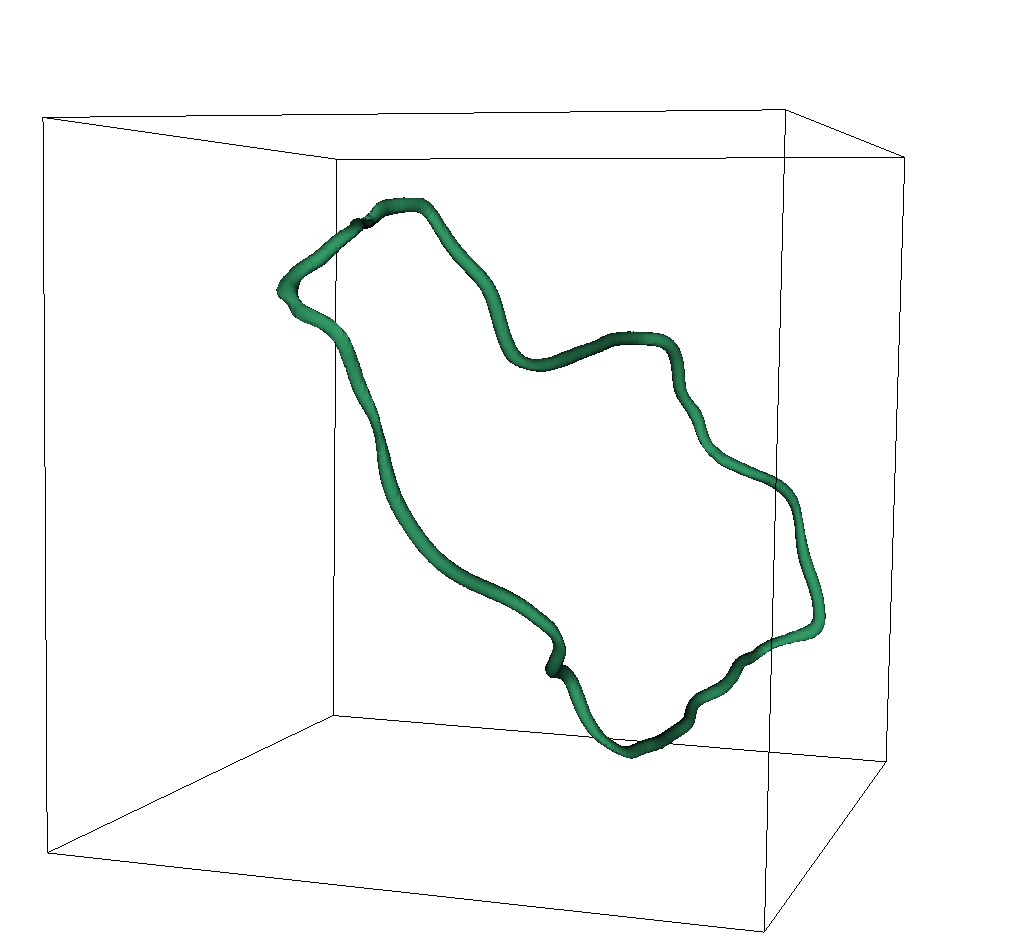}};
  \draw[->,line width=1.2mm] (-3.8,0) --(-2.65,0);
  \draw[->,line width=1.2mm] (2.2,0) --(3.35,0);
\end{tikzpicture}\vspace{0.2cm}	
 		\end{minipage}
     \caption{Three-dimensional snapshots of $|\varphi|^2=0.3v^2$ surfaces from a network simulation with $N=224$ and $\delta x =0.5/\sqrt{\lambda}v$, generated with ${\ell}_\text{str}=15/\sqrt{\lambda}v$. The snapshots correspond to the end of diffusion (left), the end of the extra-fattening phase (center), an instant when a single isolated loop is left (right).}
     \label{fig:networkevolution}
 \end{figure*}

The resulting field configuration is too energetic and contains no magnetic field. To get rid of the excess energy and allow the magnetic flux to form inside the strings, we evolve the configuration following diffusion equations of the form,
\begin{equation}\label{eq:eomfieldsdiffusion}
\begin{array}{rl}
\sqrt{\lambda}v\dot{\varphi}-\partial_i\partial_i\varphi&=-\lambda(2|\varphi|^2-v^2)\varphi\,,\\[10pt]
\sqrt{\lambda}v F_{0i}-\partial_jF_{ji} & = 2e\Im\left[\varphi^*D_i\varphi\right]\,,
\end{array}
\end{equation}
where $\dot{f}=\text{d}f/\text{d}t$. The diffusive phase is applied for $\Delta \tau_{\rm diff} = 20$ units of program time, which we find to be enough for our purposes. An illustrative example of the resulting network is presented in the left panel of fig.~\ref{fig:networkevolution}.

After the diffusion process, we let the network evolve in a radiation-dominated (RD) background, with scale factor $a(\eta)=\eta/\eta_0$, where $\eta$ indicates the conformal time and $\eta_0=70/\sqrt{\lambda} v$ in our simulations. 

While it would be possible to obtain analogous results working in Minkowski background, as done in \cite{Hindmarsh:2021mnl}, evolving the network in  RD dissipates some of the energy radiated from its decay. Moreover, we find the networks to decay slightly faster in an expanding background compared to a flat one. 

Evolving strings in an expanding background leads to a loss of resolution of the string core. To prevent this from happening, we perform an initial phase of extra-fattening~\cite{Press:1989yh}, in which the fields are evolved with equations of motion,
\begin{equation}\label{eq:eomextrafattening}
\begin{array}{rl}
\displaystyle\varphi^{\prime\prime}+2\frac{a^\prime}{a}\varphi^\prime-D_i D_i\varphi & =\displaystyle-a^{-2}\lambda \left(2|\varphi|^2-v^2\right)\varphi\,,\\[10pt]
\displaystyle\partial_0 F_{0i}+4\frac{a^\prime}{a}F_{0i}-a^{-4}\partial_j F_{ji} & \displaystyle= 2a^{-2}e\Im\left[\varphi^*D_i\varphi\right]\,,
\end{array}
\end{equation}
where $f'=\text{d}f/\text{d}\eta$. 
The extra-fattening phase is set to last for a total of  $\Delta\eta_\text{ef}=\sqrt{\eta_0(\Delta \eta_\text{HL}+\eta_0)}$, where $\Delta \eta_\text{HL}=L_{\rm B}/2$ is the {\it half-box light-crossing} time of the lattice, with $L_{\rm B}$ the side's length of the lattice {\it box}. The central panel of fig.~\ref{fig:networkevolution} shows an example of a network at the end of the extra-fattening phase. 

After this phase, the fields are evolved physically in RD, with equations of motion\\\vspace{-0.2cm}
\noindent\begin{equation}\label{eq:eomfields}
\begin{array}{rl}
\displaystyle\varphi^{\prime\prime}+2\frac{a^\prime}{a}\varphi^\prime-D_i D_i\varphi & \displaystyle=-a^2\lambda\left(2|\varphi|^2-v^2\right)\varphi\,,\\[10pt]
\displaystyle\partial_0 F_{0i}-\partial_j F_{ji} & \displaystyle= 2a^2e\Im[\varphi^*D_i\varphi]\,,
\end{array}
\end{equation}
for an additional time $\Delta \eta_\text{RD}=\Delta \eta_\text{HL}-\Delta\eta_\text{ef}$. This procedure ensures that, by the end of the physical evolution phase, the width of the string is equal to that at the end of diffusion.

When the evolution in a RD background ends, the string network is close to the scaling regime. 
In most cases, however, the network has not yet decayed into a single loop. We subsequently evolve the resulting network in a Minkowski background so that it is able to decay. 
Note that, when changing from RD to Minkowski, it takes some time for the network to adapt to the new background, since the characteristics of the scaling regime depend on the background metric.  After changing the metric, we wait for half a half-box light-crossing time, $\Delta t_\text{HL}/2$, before starting to study any isolated loop that may arise. We call this period a \textit{transient phase}. We believe this minimizes the impact of the sudden change of background. 
The evolution in Minkowski background is held for a maximum time of $2\Delta t_\text{HL}$, including the transient phase, which we found enough to typically have one isolated string loop remaining. An example of an isolated network is presented in the right panel of fig.~\ref{fig:networkevolution}.

After the transient period, if an isolated loop is found we turn on the emission of GWs, and study the evolution of the loop until it disappears. Approximately, $\sim 20\%$ of our simulations lead to isolated loops, of which $\sim 80\%$ can be used for our study. We discard those loops that self intersect forming several loops of similar size or infinite strings. Altogether, only $\sim16\%$ of the simulations are suitable for this work. \\

\subsection{Artificial loops of type I}

Artificial loops of type I are generated from the intersection of two pairs of parallel infinite boosted strings, following the procedure used in \cite{Matsunami:2019fss}---see also \cite{Saurabh:2020pqe,Baeza-Ballesteros:2023say}. We choose the initial configuration so that one of the two loops resulting from the intersection of the infinite strings is much larger than the other one, and wait for the smaller one to decay before starting our study of the longer loop.  

We now describe the initialization procedure in detail. We consider one pair parallel to the $z$ axis and the other parallel to the $x$ axis, and we refer to each of them with subscripts ``1'' and ``2'', respectively, which should not be confused with the component index of the gauge field, $\mu=0,1,2,3$.

We first explain how the pair of strings parallel to the $z$ axis is generated. The starting point is the solution for the Nielsen-Olsen (NO) vortex \cite{Nielsen:1973cs} in the temporal gauge, $\varphi^{(k)}_\text{NO}$ and $A^{(k)}_{\text{NO},\mu}$, where $k=\pm 1$ indicates the winding number of the string. The static NO configuration is boosted in the $(x,y)$-plane with velocity $\vec{v}_1=v_1(\sin\alpha_1,\cos\alpha_1)$, resulting in
\begin{widetext}
\begin{equation}
\begin{array}{rl}
\bar{\varphi}_{\vec{v}_1}^{(\pm)}(x,y;t)&=\varphi_\text{NO}^{(\pm)}(x^\prime,y^\prime)\,,\\[8pt]
\bar{A}_{\vec{v}_1,0}^{(\pm)}(x,y;t)&=-\gamma_1 s_1 v_1 A_{\text{NO},1}^{(\pm)}(x^\prime,y^\prime)-\gamma_1 c_1 v_1 A_{\text{NO},2}^{(\pm)}(x^\prime,y^\prime)\,,\\[8pt]
\bar{A}_{\vec{v}_1,1}^{(\pm)}(x,y;t)&=[1+(\gamma_1-1)s_1^2] A_{\text{NO},1}^{(\pm)}(x^\prime,y^\prime)+(\gamma_1-1)s_1c_1 A_{\text{NO},2}^{(\pm)}(x^\prime,y^\prime)\,,\\[8pt]
\bar{A}_{\vec{v}_1,2}^{(\pm)}(x,y;t)&=(\gamma_1-1)s_1c_1 A_{\text{NO},1}^{(\pm)}(x^\prime,y^\prime)+[1+(\gamma_1-1)c_1^2] A_{\text{NO},2}^{(\pm)}(x^\prime,y^\prime)\,,
\end{array}
\end{equation}
\end{widetext}
with $s_1=\sin\alpha_1$, $c_1=\cos\alpha_1$ and $\gamma_1=(1-v_1^2)^{1/2}$. Here $(x^\prime,y^\prime)$ are the coordinates in the rest frame of the string and $(t,x,y)$ are the coordinates in the boosted frame, related by
\begin{equation}\label{eq:relativisticboost}
\begin{array}{rl}
   x'&=-\gamma_1 v_1 s_1 t + [1+(\gamma_1-1)s_1^2] x+(\gamma_1-1)s_1c_1y\,,\\[5pt]
   y'&= -\gamma_1 v_1 c_1 t + (\gamma_1-1)s_1c_1x+[1+(\gamma_1-1)c_1^2]y\,.
\end{array}
\end{equation}

The relativistic boost produces an undesired time component of the gauge field. To go back to the temporal gauge, we perform a gauge transformation, 
\begin{equation}
\varphi = \text{e}^{ie\xi}\bar{\varphi}\,,\quad\quad\quad A_\mu=\bar{A}_\mu-\partial_\mu\xi\,,
\end{equation}
where $\xi\equiv\xi(x,y)$ is a function chosen so that $A_0=0$ in the boosted frame,
\begin{equation}
\dot{\xi} = \bar{A}_0\longrightarrow \xi=\int_0^t A_0\text{d} t\,.
\end{equation}
As we evaluate the initial configuration at $t=0$, we can set $\xi=0$. However, $\dot{\xi}=\bar{A}_0$, which we need to take into account to compute the time derivatives of the fields,
\begin{equation}\label{eq:gaugetransformationderivative}
\dot{\varphi}=\dot{\bar{\varphi}}-ie\bar{A}_0\bar{\varphi}\,,\quad\quad\quad
\dot{A}_i=\dot{\bar{A}}_i-\partial_i\bar{A}_0\,.
\end{equation}

The product ansatz can then be used to generate a pair of parallel boosted strings. The complex fields for both strings, evaluated at $t=0$, are multiplied, while the gauge fields are summed,
\begin{widetext}
\begin{equation}\label{eq:productansatzsinglepair}
\begin{array}{rl}
    \varphi_1(x,y;t)&=\displaystyle\frac{1}{v}\varphi_{{\bm v}_1}^{(+)}\left[x-\left(\frac{L_{\rm B}}{2}+a_1\right),y-\left(\frac{L_{\rm B}}{2}+b_1\right);t\right] 
     \displaystyle\cdot \varphi_{-{\bm v}_1}^{(-)}\left[x-\left(\frac{L_{\rm B}}{2}-a_1\right),y-\left(\frac{L_{\rm B}}{2}-b_1\right);t\right]\,,\\[13pt]
    A_{1, \mu}(x,y;t) &=\displaystyle A_{{\bm v}_1,\mu}^{(+)}\left[x-\left(\frac{L_{\rm B}}{2}+a_1\right),y-\left(\frac{L_{\rm B}}{2}+b_1\right);t\right] 
     \displaystyle + A_{-{\bm v}_1,\mu}^{(-)}\left[x-\left(\frac{L_{\rm B}}{2}-a_1\right),y-\left(\frac{L_{\rm B}}{2}-b_1\right);t\right]\,,
     \end{array}
\end{equation}
\end{widetext}
where $a_1$ and $b_1$ indicate the distance of each string to the center of the lattice. 
The corresponding time derivatives are straightforward to evaluate by successive application of the chain rule. 

The resulting configuration is then modified to fit in a periodic lattice, using a similar approach as in \cite{Matsunami:2019fss}. We do not modify the gauge field, as we find its long-distance energy contribution to be negligible. The scalar field approaches the vacuum exponentially fast far from the string, and so we only need to change its phase,
\begin{equation}\label{eq:hperiodicitymodification}
\varphi_1=|\varphi_1|\text{e}^{i\theta_1}\longrightarrow \varphi_1^\text{per}=|\varphi_1|\text{e}^{ih(x,y)\theta_1}\,,
\end{equation}
which affects the time-derivative of the field. The filter function $h(x,y)$ is chosen so that the phase changes smoothly close to the boundary towards zero. We opt to use
\begin{equation}\label{eq:filterfunction}
h(x,y)=\left\{
\begin{array}{ll}
\displaystyle\frac{L_{\rm B}/2-|x_L|}{L_{\rm B}/2-L_h}\,, &\quad |x_L|> L_h\,,\,\,|x_L|\geq|y_L|\,,\\[10pt]
\displaystyle\frac{L_{\rm B}/2-|y_L|}{L_{\rm B}/2-L_h}\,, &\quad |y_L|> L_h\,,\,\,|x_L|<|y_L|\,,\\[10pt]
1\,, & \quad\text{otherwise}\,,
\end{array}
\right.
\end{equation} 
where we use the shorthand notation $x_L=x-L_{\rm B}/2$ and $y_L=y-L_{\rm B}/2$.
This differs from the choice in \cite{Matsunami:2019fss}, which we have found leaves some residual energy close to the $(x=0,L_{\rm B},y=L_{\rm B}/2)$ boundaries that leads to instabilities at late times in the simulations. We use $L_h = L_{\rm B}/2-16/\sqrt{\lambda}v$ in our simulations, independently of the size of the lattice. We have checked that varying $L_h$ up to a factor of four has a negligible effect on the final results. 

\begin{figure*}[t]
\centering
\begin{minipage}{1\textwidth}
      \centering
 		\begin{tikzpicture}[>={LaTeX[width=3mm,length=4mm]},->]
  \node[] at (-6.,0) {\includegraphics[width=0.29\textwidth]{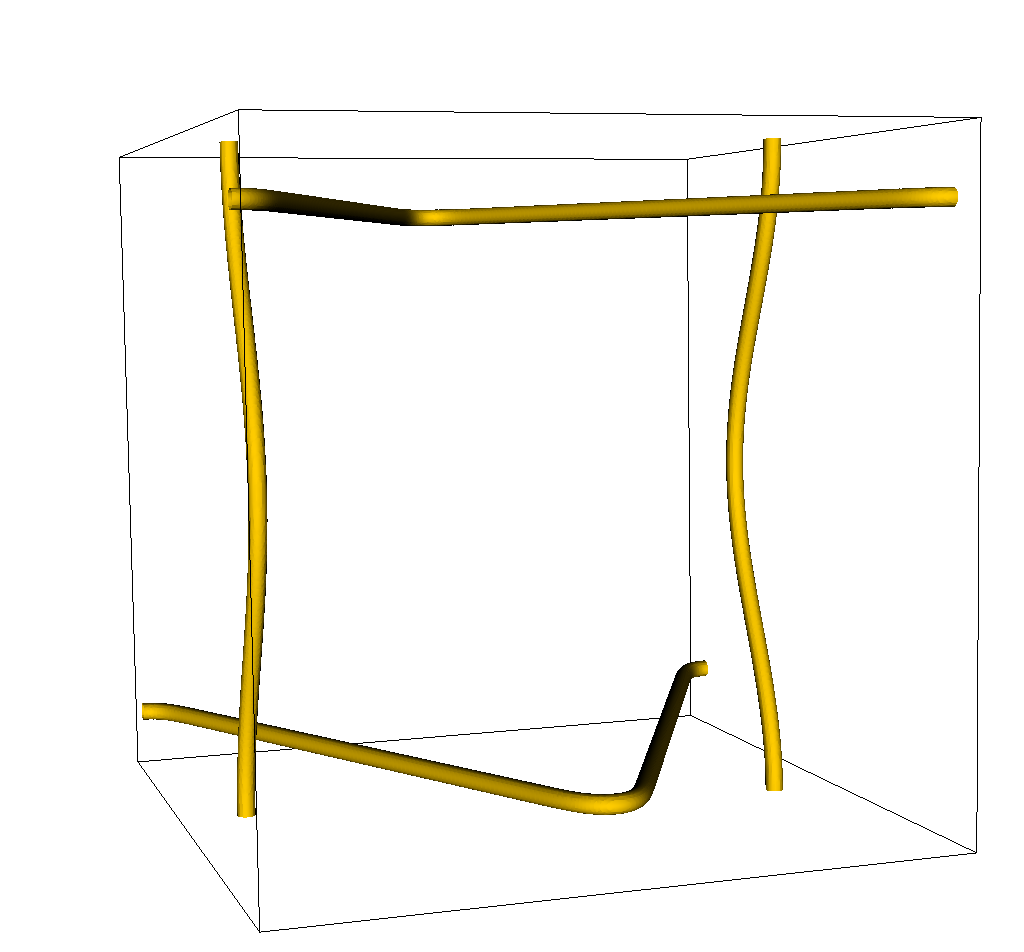}}; 
  \node[] at (-1,0) {\includegraphics[width=0.29\textwidth]{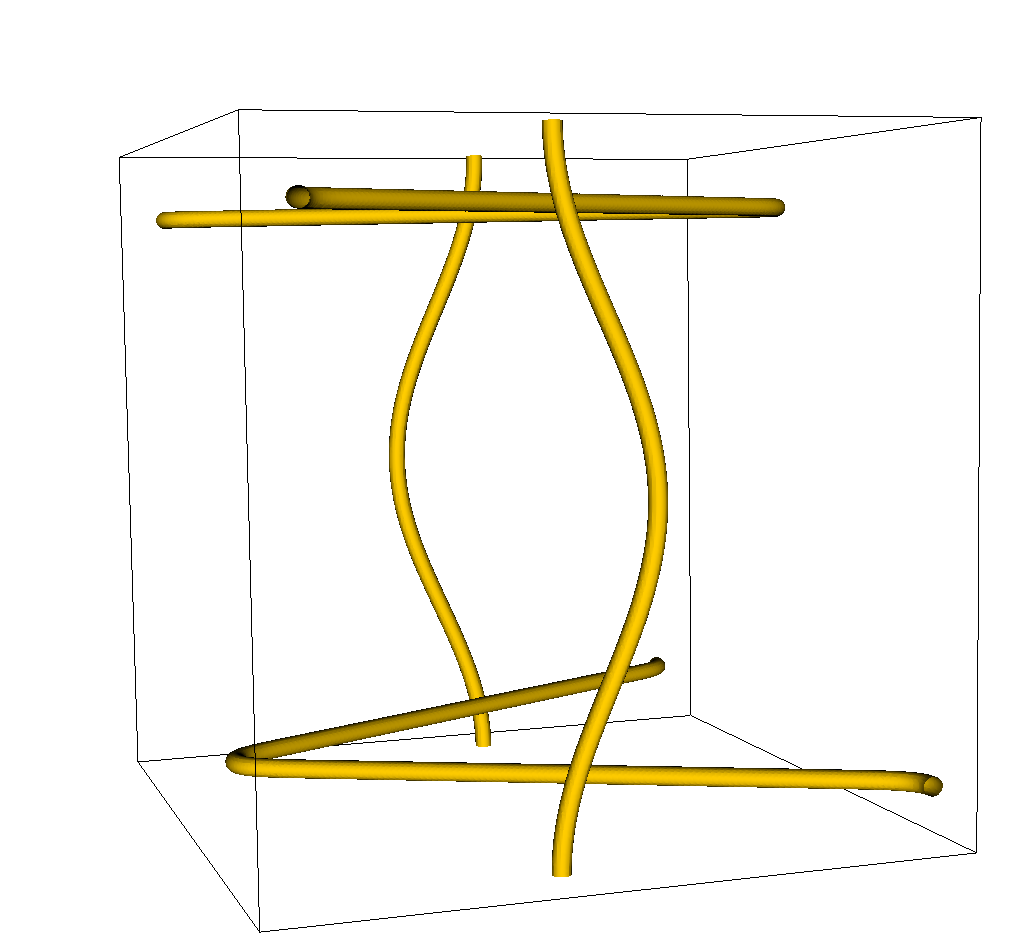}};
  \node[] at (6.,0) {\includegraphics[width=0.29\textwidth]{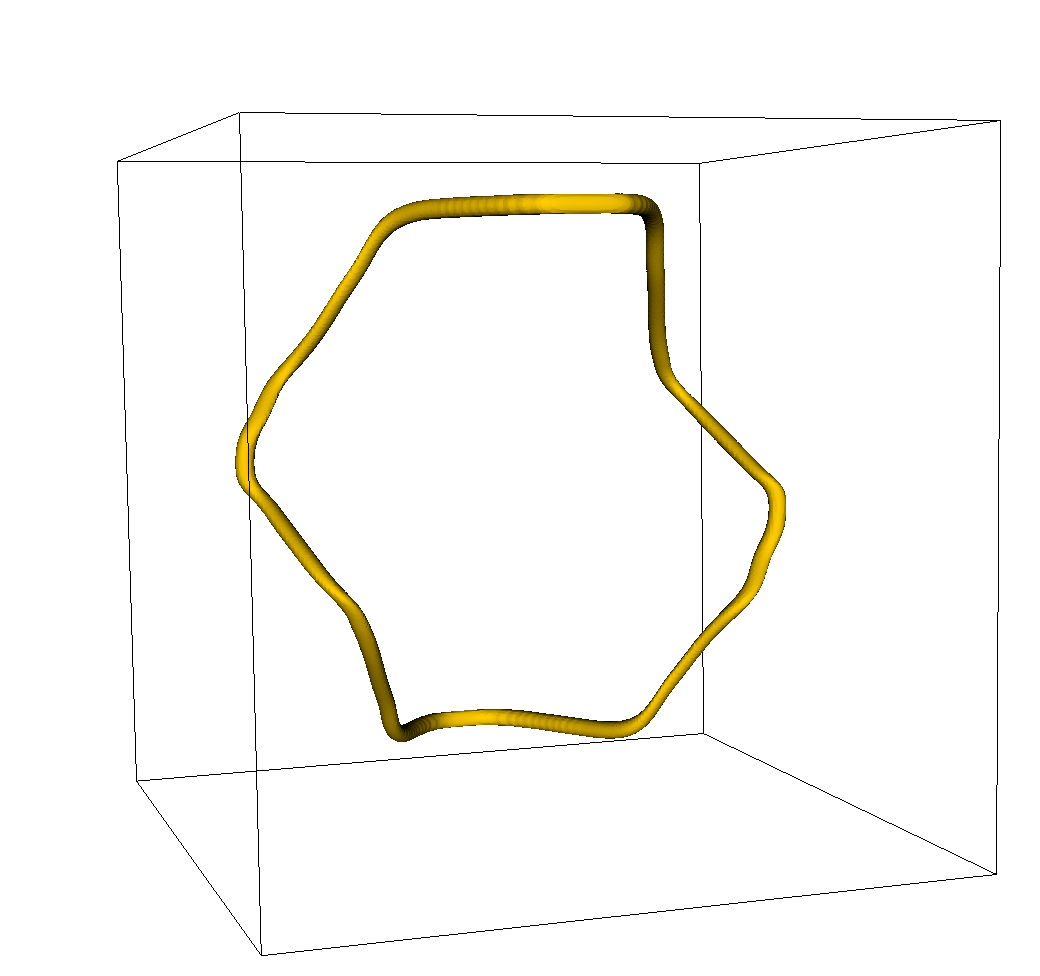}};
  \draw[->,line width=1.2mm] (1.9,0) --(3.5,0);
\end{tikzpicture}\vspace{0.cm}	
 		\end{minipage}
    \caption{
        Three-dimensional snapshots of $|\varphi|^2=0.2v^2$ surfaces of the simulation of an artificial loop of type II, corresponding to the end of diffusion (left and center) and the moment in which a single loop is produced (right).}
    \label{fig:windingloopisolated}
\end{figure*} 

Finally, we use again the product ansatz on two perpendicular string pairs and generate the initial conditions for our simulations,
\begin{equation}
\begin{array}{rl}
\varphi(x,y,z) &= \varphi_1^\text{per}(x,y;t=0) \cdot \varphi_2^\text{per}(z,y;t=0)\,,\\[5pt]  
A_\mu(x,y,z) &= A_{1,\mu}(x,y;t=0) + A_{2,\mu}(z,y;t=0)\,.  
\end{array}
\end{equation}
The time derivatives of the fields are computed by successive differentiation, taking into account the gauge transformations in \cref{eq:gaugetransformationderivative} and the use of the filter function in \cref{eq:hperiodicitymodification}.

In this work, we consider different boost velocities for the two pairs, $v_1\neq v_2$ and set $\alpha_1=-\alpha_2=\alpha$, as we observe this leads to longer-lived strings. More concretely, we have found that other choices of the boost direction, such as $\alpha_1=\alpha_2$,  lead to a rapid double-line collapse (dLC). This is a physical field-theory phenomenon happening when two parallel segments of string approach each other completely annihilating. However, we believe that its occurrence is a result of the artificial square initial configuration, and so we choose the initial conditions to prevent it from happening. In those cases in which dLC still occurs, we use a fit to factor out its effect from the emission power of particles, as explained in \cref{sec:resultsdecay}.

The initial configuration is evolved in a Minkowski background and the four strings soon intersect forming two loops. We consider $b_1,b_2\ll L$, so that the strings intersect rapidly after the simulation is started, and choose $a_1=a_2$ small compared to the box size, so that the inner loop is much smaller than the outer one collapses rapidly after the start of the simulation. After this happens, we turn on GWs and analyze the evolution of the fields until some time after the loops collapses. We note that no isolation procedure is used to separate the loops, as was done in \cite{Baeza-Ballesteros:2023say}. We have found that a naive generalization of the technique proposed there for global strings breaks Gauss' law, leading to unstable simulations.

Furthermore, choosing $a_1=a_2$ small ensures the initial infinite strings are far from the region modified by the filter in \cref{eq:filterfunction}. Some parts of the strings of each pair still lies on top of the modified region of the opposite pair. The size of this region depends on the choice of $L_h$ and, as discussed above, we observe no effect on the dynamics of the loops from changing this parameter. Thus, we believe the effect of the chosen filter function on the loop dynamics to be negligible. 

\subsection{Artificial loops of type II}

Artificial loops of type II are generated following the procedure introduced in \cite{Hindmarsh:2017qff,Hindmarsh:2021mnl}. This is based on initializing static strings of arbitrary shape by setting some magnetic flux on the plaquettes pierced by the initial strings. We note that this techniques relies on the use of the hybrid~\cite{Bevis:2006mj,Hindmarsh:2017qff} or the compact~\cite{Figueroa:2020rrl} formulation of the gauge theory on the lattice, in which the field-strength tensor is discretized in terms of the link variables. For our study, we use the hybrid formulation.

Given some shape of the desired one-dimensional strings, one can determine a set of plaquettes pierced by the string. The idea is to set the magnetic flux through these plaquettes to $\pm2\pi$, depending on the direction from which the plaquette is pierced by the string. This is achieved by setting the gauge variables on the links to
\begin{equation}
A_\mu(\bm{n})=\pm\frac{\pi}{2e\delta x }\,,
\end{equation}
where the sign depends on the orientation of link. The complex scalar field is set to be equal to the vacuum expectation value, $\varphi=v$, everywhere. 
The initial configuration is then diffused for five units of program time, using \cref{eq:eomfieldsdiffusion}, which leads to the formation of strings with the expected radius, $r_\text{c}\sim m_\text{s}^{-1}$. 

For our study, we use an initial configuration composed by four non-straight static strings, following \cite{Hindmarsh:2021mnl}, which intersect soon after the start of the simulation forming two loops. We consider two strings at fixed $y=L_{\rm B}/10$ and $y=9L_{\rm B}/10$ with sinusoidal form. The coordinates of the string core are given by
\begin{equation}
x=\pm A \cos(2\pi z/L_{\rm B})\,,
\end{equation}
with each sign corresponding to a different value of $y$. Also, we set $A=0.075L_{\rm B}$. The other two strings have fixed $z$ with a {\it sawtooth} form,
\begin{equation}
x=\left\{\begin{array}{ll}
\displaystyle\pm C\left[\frac{y}{L_{\rm B}/4}-1\right]\,,\quad\quad\quad & 0 \leq y \leq L_{\rm B}/2\,,\\[10pt]
\displaystyle\mp C\left[\frac{y}{L_{\rm B}/4}-3\right]\,,\quad\quad\quad & L_{\rm B}/2 < y < L_{\rm B}\,,
\end{array}\right.
\end{equation}
with $z=L_{\rm B}/10$ and $z=9L_{\rm B}/10$, for each string, and $C=L_{\rm B}/2$. The signs, again, corresponds to each possible value of $z$. A representation of the resulting configuration at the end the diffusion period, is shown in the left and central panels of \cref{fig:windingloopisolated}, from two different perspectives.

After the diffusive phase, the strings are let to evolve in Minkoski spacetime, using a discretization of the equations of motion consistent with the hybrid formulation. The four infinite strings start to move as a result of their non-straight form, and they eventually intersect, forming two loops. Due to the initial position of the infinite strings, the outer one is smaller than the inner one. We wait until the outer loop disappears and study the inner one afterwards. An example of the resulting isolated loop is shown in the right panel of \cref{fig:windingloopisolated}.

\begin{figure*}[t]
\includegraphics[width=1\textwidth,height=6.3cm]{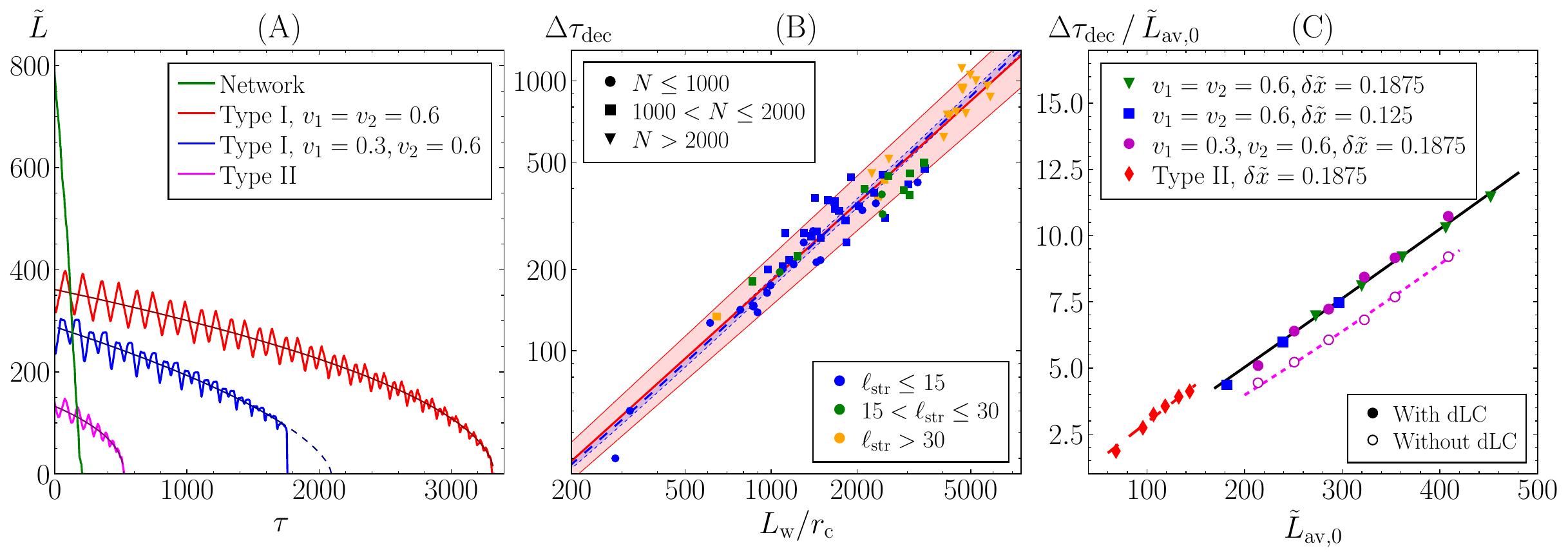}
\caption{Panel (A): Evolution of $\tilde L$ of a network loop (green), artificial loops of type I (red and blue) and type II (magenta). Here $\tau=0$ signals the moment when an isolated loop is left in the simulation. Solid lines 
for artificial loops 
represent a fit to an oscillation-averaged length, $\tilde L_{\rm av}$, and the dashed line an extrapolation of the previous fit to $\tau_\text{dec}$. 
Panel (B): $\Delta \tau_{\rm dec}$ vs $\tilde{L}_0$ for network loops, varying $N$ and $\ell_{\rm str}$, with blue and red shaded regions corresponding to linear and power-law fits, respectively. Panel (C): $\Delta \tau_{\rm dec}/\tilde{L}_\text{av,0}$ vs $\tilde{L}_\text{av,0}$ for artificial loops, with dashed, solid and dotted lines, representing power-law fits.}
\label{fig:stringdynamics}
\end{figure*}

A drawback of this initialization method is that there remains a magnetic flux frozen on the  initialized \mbox{plaquettes}. When measuring the winding number on these plaquettes one finds a non-zero results, which can be associated to a \textit{Dirac string}~\cite{Hindmarsh:2017qff}. This is an unphysical consequence of this particular initialization procedure, since these Dirac strings do not contain energy, and have not been observed to affect the dynamics of the physical strings. When measuring the length of artificial strings of type II from the number of pierced plaquettes, we subtract the number of initialized plaquettes to the count. For our loops, the number of plaquettes belonging simultaneously  to both the real and the ghost string is negligible compared to the total number of plaquettes, and so this does not affect our ability to measure the length of the loops.

Overall, artificial loops of type II are found to behave very similarly to loops of type I. However, loops of type II are typically much smaller than those of type I generated in lattices of the same size. Thus, we consider type II loops to study the power emission of particles, but restrict our investigation on GW emission to type I loops, for simplicity.


\section{Results on loop decay} \label{sec:resultsdecay}

We have studied 
69 network loops, 14 artificial loops of type I, and 6 of type II, with initial length-to-core width ratios $200 \lesssim L_0/r_\text{c}\lesssim 6000$ (network) and $120\lesssim L_0/r_\text{c}\lesssim 640$ (artificial). We have characterized the dependence of the decay time of 
loops $\Delta \tau_\text{dec}$, 
on their initial length $\tilde L_0$, and energy $\tilde E_\text{str,0}$. For details on the simulations, see \cref{app:simulationdetails}.

\begin{figure*}[!t]
    \centering
\begin{minipage}{0.19\textwidth}
	\includegraphics[width=\textwidth]{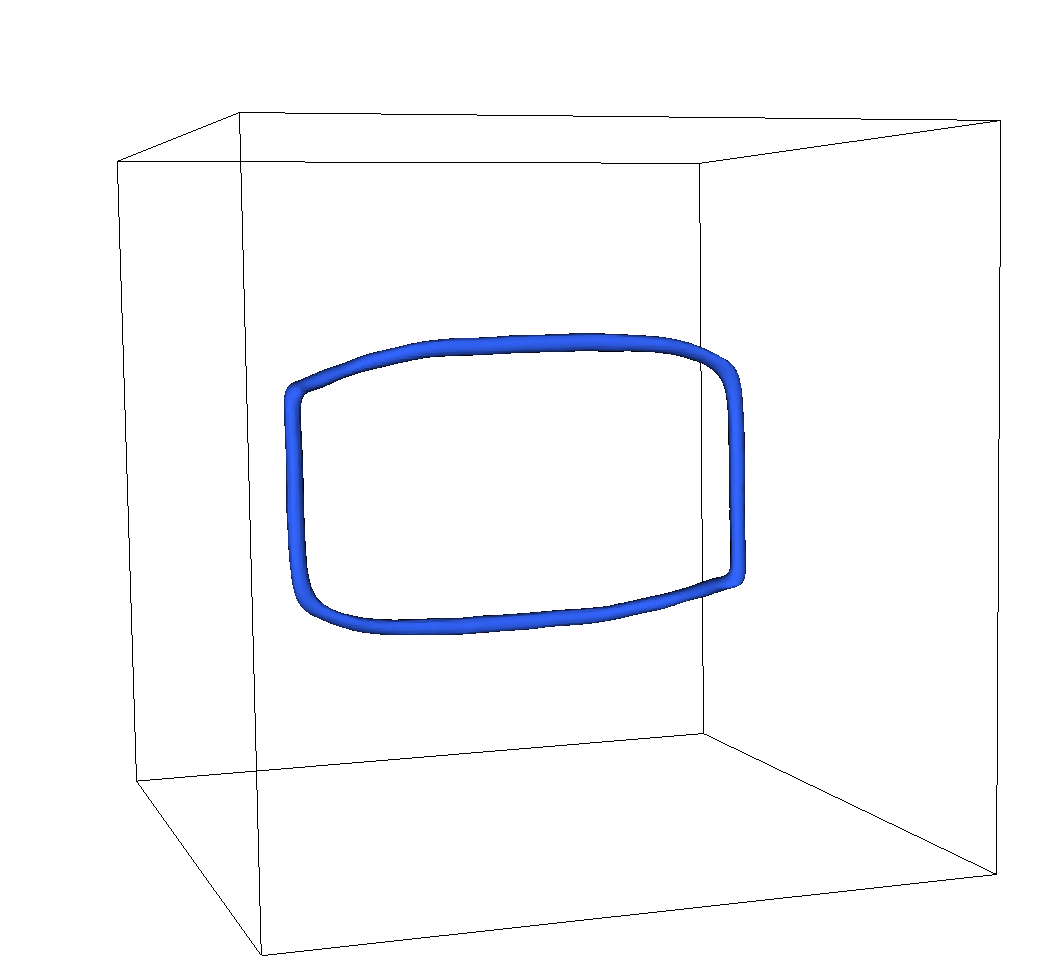}
\end{minipage}
\begin{minipage}{0.19\textwidth}
	\includegraphics[width=\textwidth]{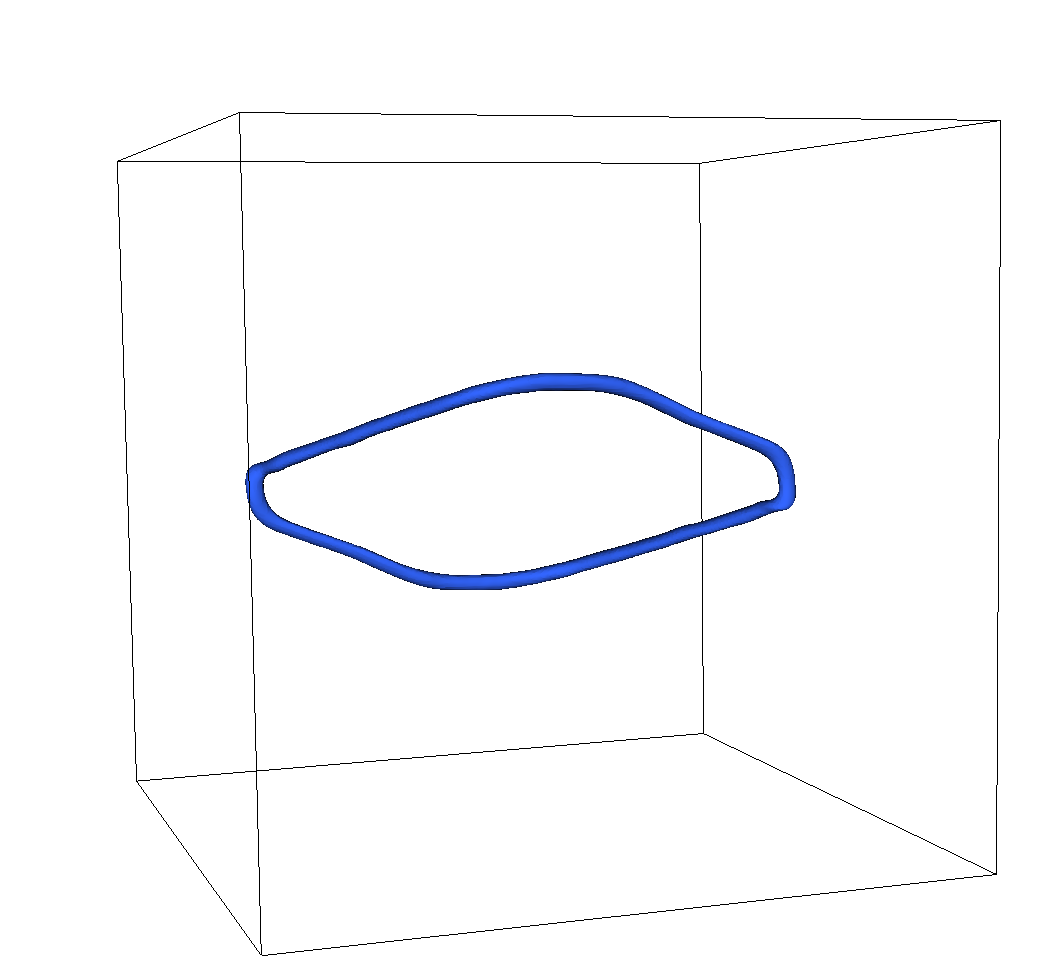}
\end{minipage}
\begin{minipage}{0.19\textwidth}
	\includegraphics[width=\textwidth]{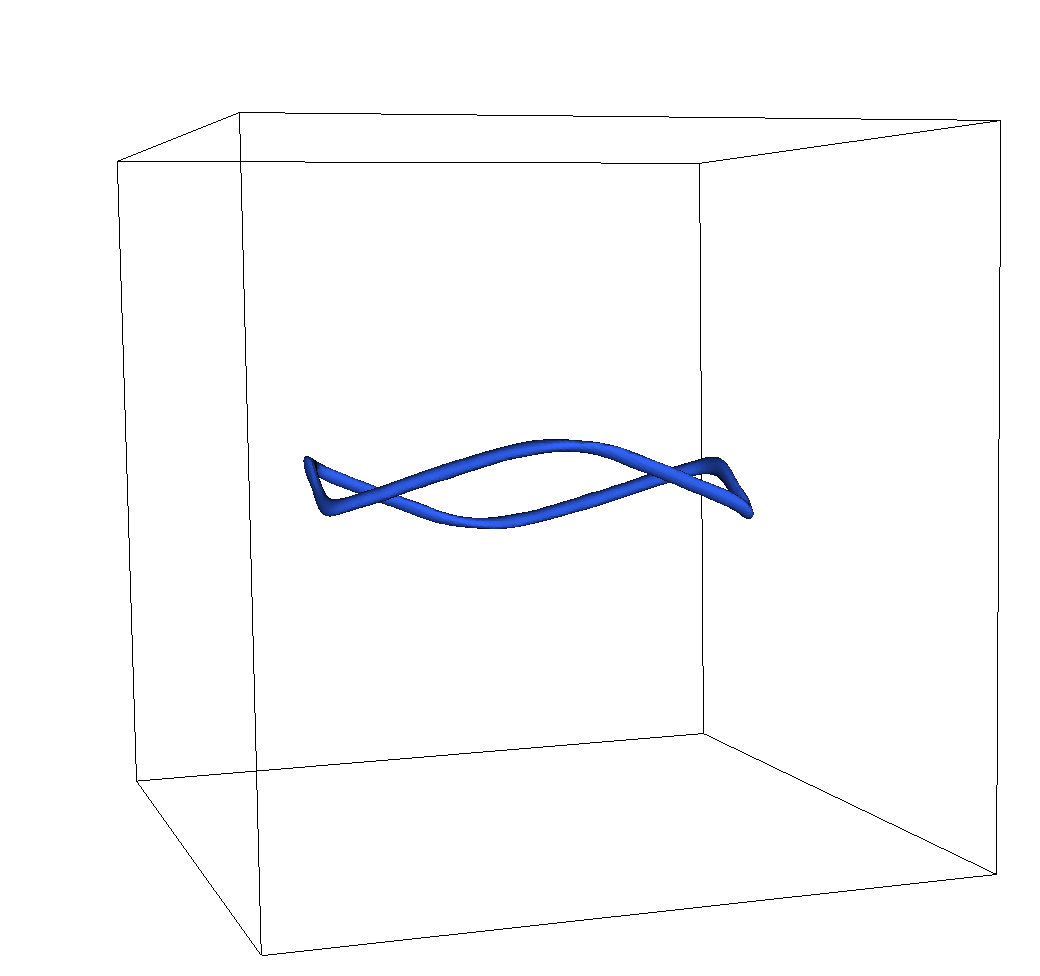}
\end{minipage}
\begin{minipage}{0.19\textwidth}
	\includegraphics[width=\textwidth]{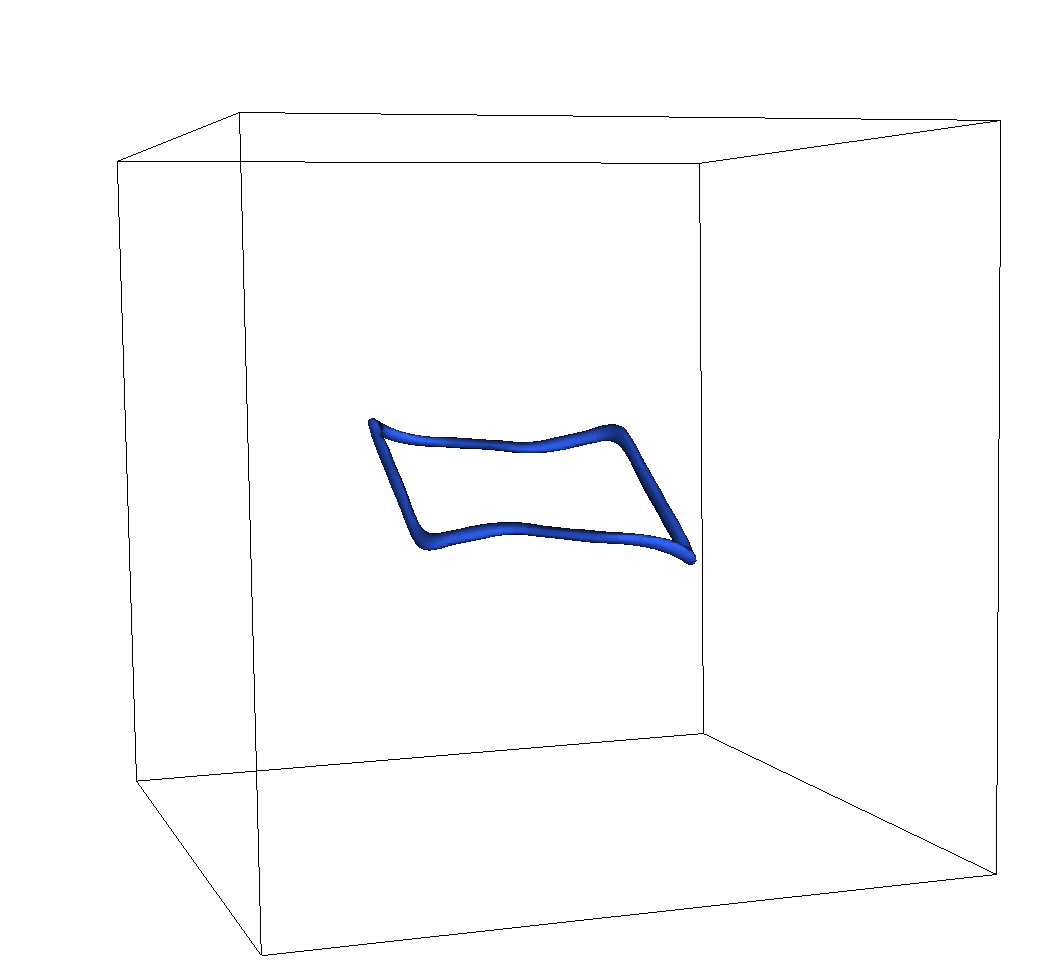}
\end{minipage}
\begin{minipage}{0.19\textwidth}
	\includegraphics[width=\textwidth]{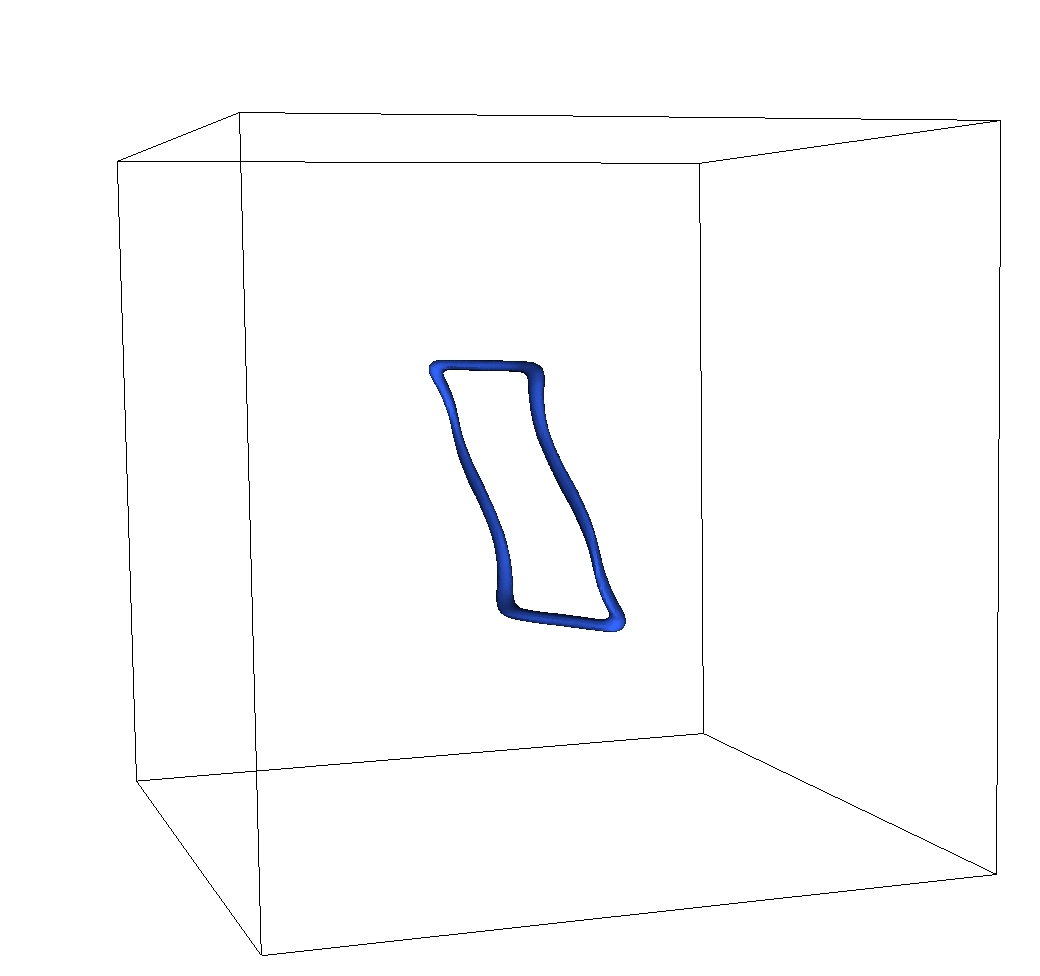}
\end{minipage}
\vspace{0.2cm}

\centering	
\begin{minipage}{0.19\textwidth}
	\includegraphics[width=\textwidth]{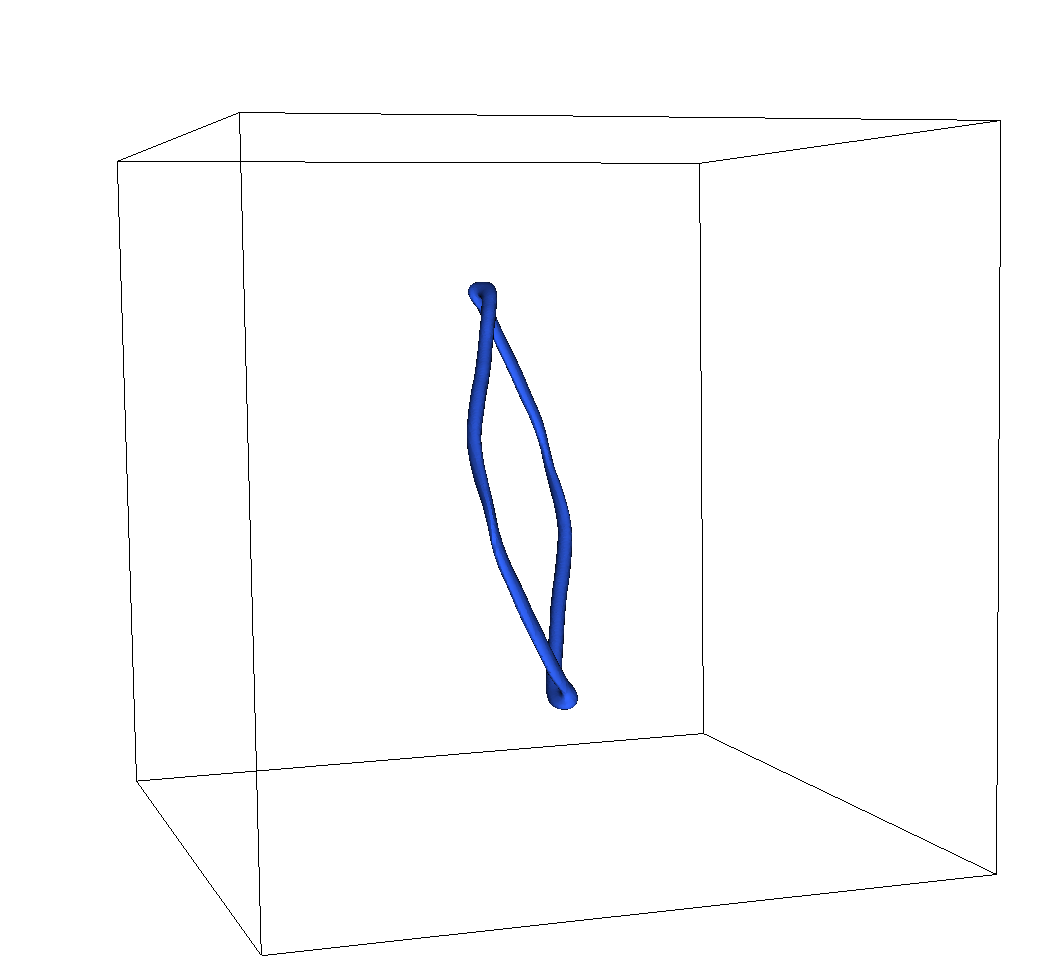}
\end{minipage}
\begin{minipage}{0.19\textwidth}
	\includegraphics[width=\textwidth]{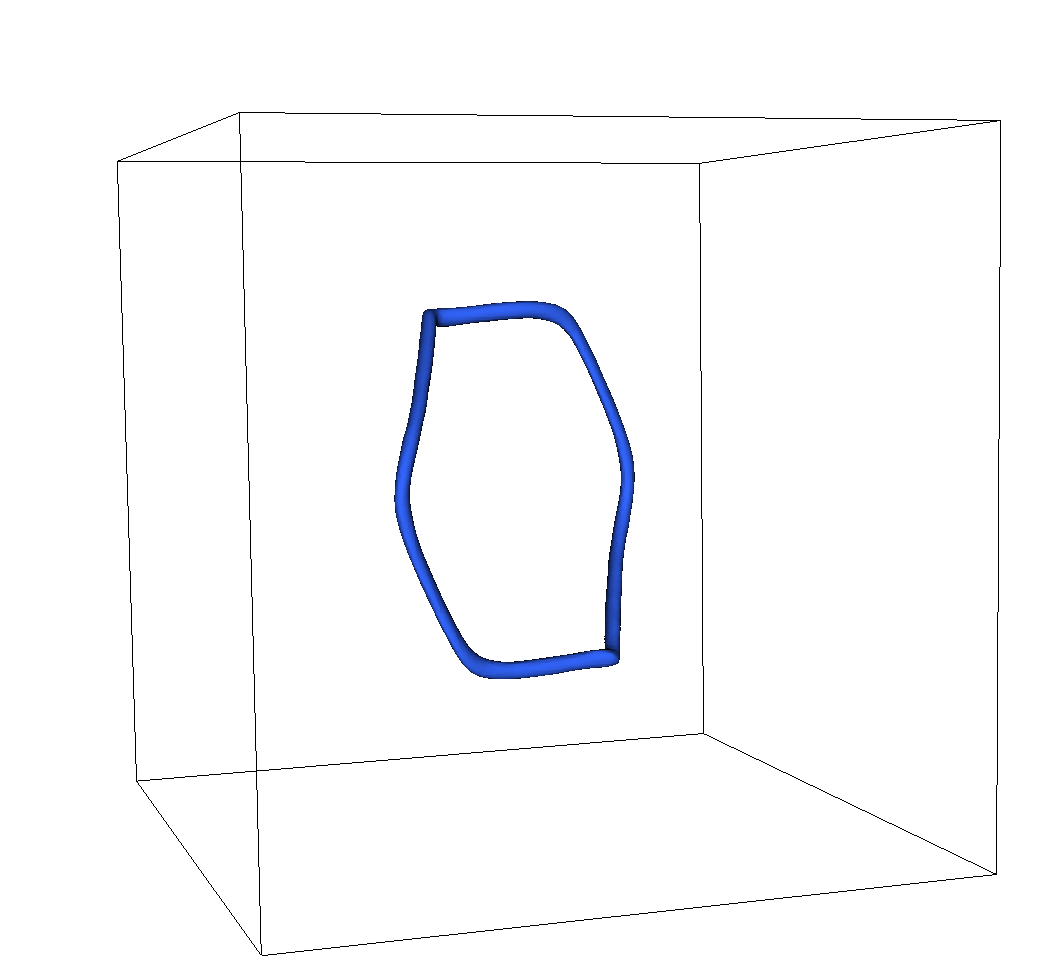}
\end{minipage}
\begin{minipage}{0.19\textwidth}
	\includegraphics[width=\textwidth]{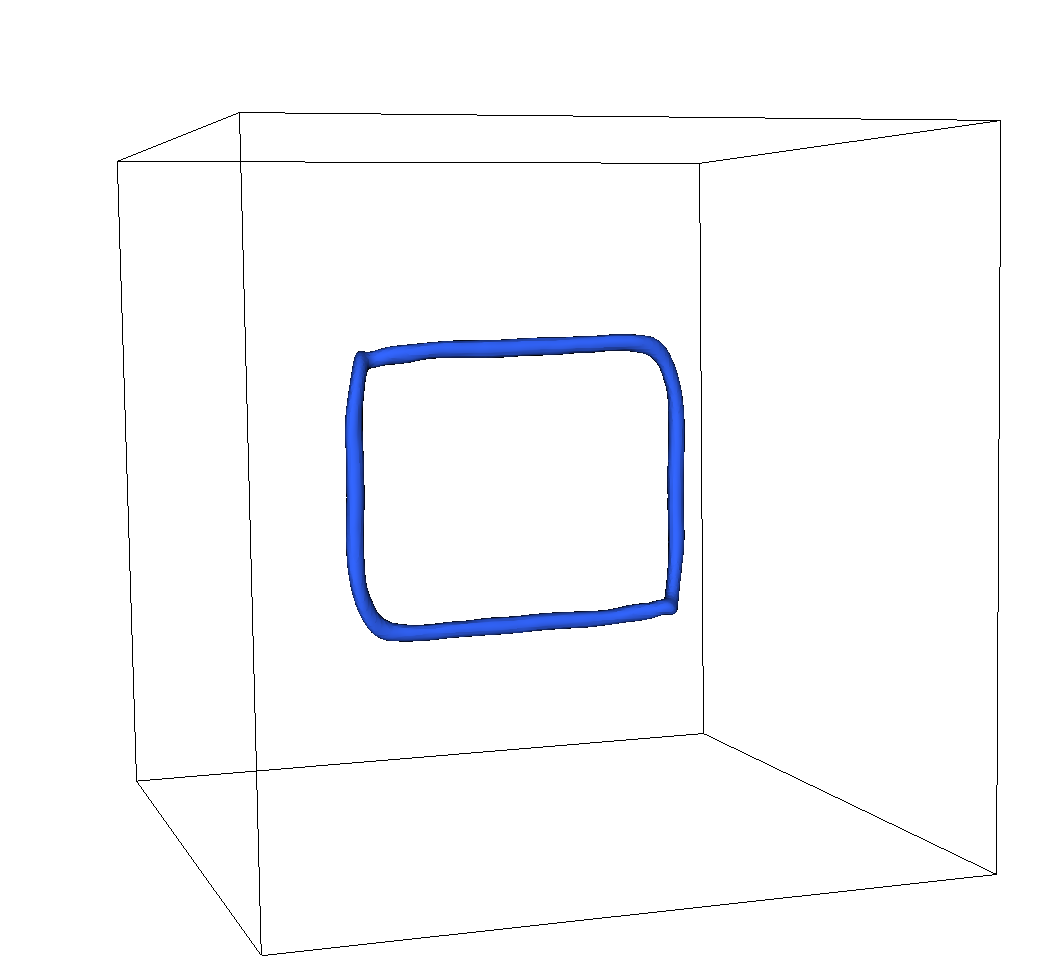}
\end{minipage}
\begin{minipage}{0.19\textwidth}
	\includegraphics[width=\textwidth]{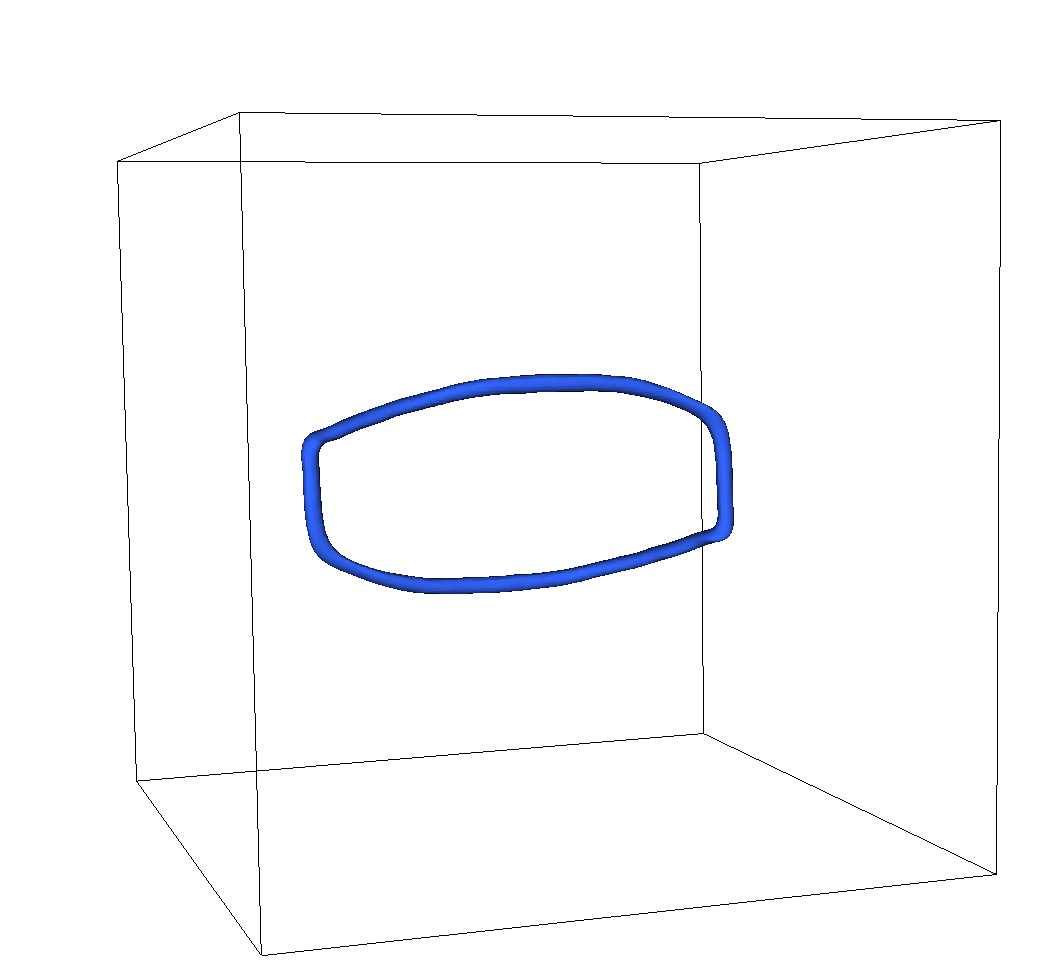}
\end{minipage}
\begin{minipage}{0.19\textwidth}
	\includegraphics[width=\textwidth]{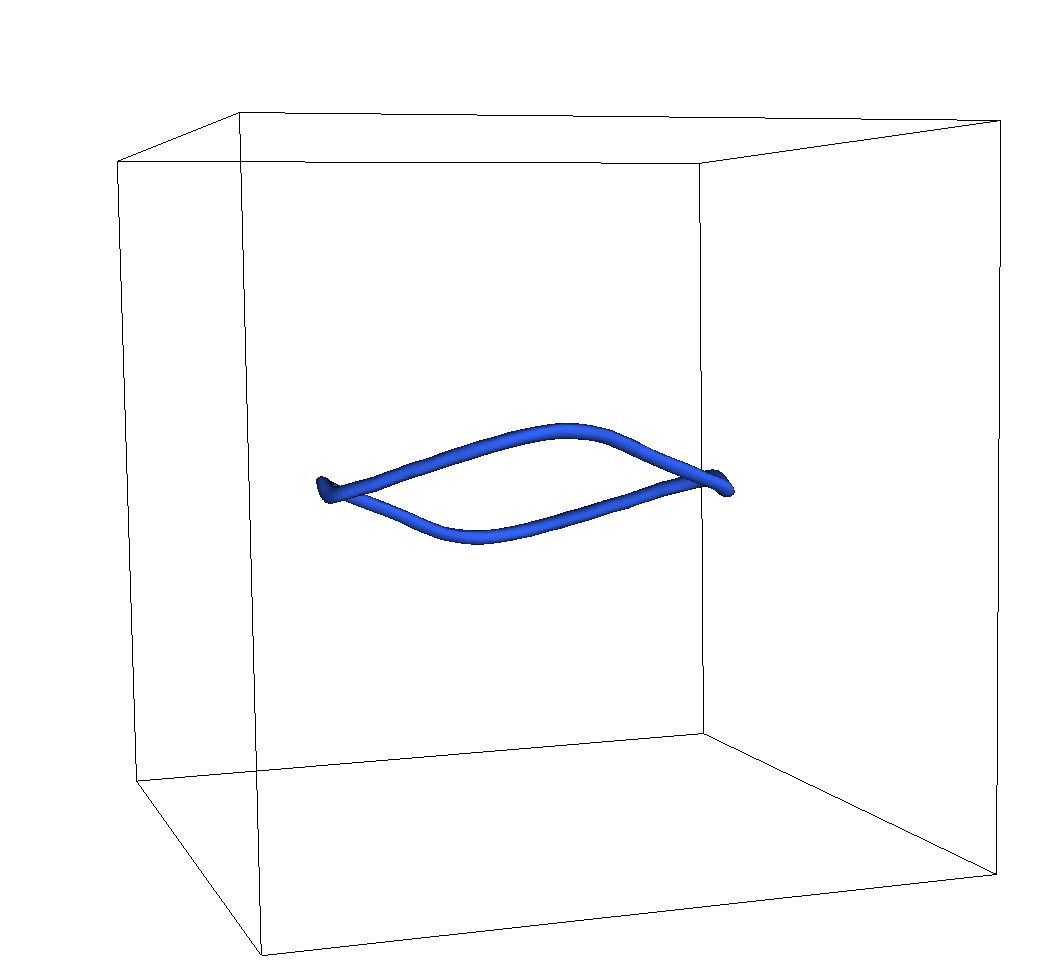}
\end{minipage}

    \caption{
        Three-dimensional snapshots of $|\varphi|^2=0.2v^2$ surfaces of an oscillating artificial loop of type I, simulated with $N=256$, $\delta\tilde{x}=0.25$, $v_1=v_2=0.25$ and $\sin\alpha=0.4$. Time goes from left to right and from top to bottom, with snapshots separated by $\Delta t=10/\sqrt{\lambda}v$ units of program time. The field has been periodically shifted by $L_{\rm B}/2$ in the $x$ and $z$ directions for clarity in the representation.}
    \label{fig:stringsnapshotsartificial}
\end{figure*} 

Fig.~\ref{fig:stringdynamics}-(A) shows the length evolution of 
representative loops. 
The length of a network loop (in green) decays 
rapidly in time, with no oscillations. 
Artificial loops, however, live much longer and oscillate many times before decaying--see fig.~\ref{fig:stringsnapshotsartificial} for an example of the evolution of an artificial loop of type I over one full oscillation. 
In the case of type I loops, depending on the boost velocities ($v_1,\,v_2$), the loop either disappears smoothly (in red), or it disappears 
abruptly (in blue) due to a ``double-line collapse" (dLC), when two parallel segments approach each other and annihilate.
Type II loops display a similar decay pattern (in magenta). 
As dLC is a result of the artificial initial condition, we remove its effect on our results for artificial loops by performing a 3-parameter fit to an oscillation-averaged length $\tilde{L}_\text{av}(\tau)=C(\tau_\text{dec}-\tau)^p$~\cite{Hindmarsh:2021mnl} (solid lines, with $p \approx 1/2$). This allows  to determine: 
$i)$ $\Delta\tau_\text{dec}=\tau_\text{dec}-\tau_0$ as the lifetime, with $\tau_0$ the time when the loop becomes isolated 
in the simulation, and $ii)$ 
the initial (oscillation-averaged) length of the isolated loop, $\tilde L_{{\rm av},0}=\tilde L_\text{av}(\tau_0)$.  
A similar behaviour is observed fitting against the loops' energy as $\tilde{E}_\text{av}=D(\tau_\text{dec}-\tau)^q$, with $q \approx 1/2$, leading to an estimate of the (oscillation-averaged) initial string energy $\tilde E_\text{av,0}$.

Fig.~\ref{fig:stringdynamics}-(B) shows $\Delta \tau_{\rm dec}$ vs $\tilde L_0$ for all network loops studied. As in~\cite{Hindmarsh:2021mnl,Baeza-Ballesteros:2023say}, we observe $\Delta \tau_{\rm dec}$ roughly scaling linearly with $\tilde L_0$. We quantify this by performing two different fits, a linear fit  $\Delta\tau_\text{\rm dec}=c_1\tilde{L}_0+c_2$ (blue dashed line and band), and a power-law fit $\Delta\tau_{\rm dec}=A\tilde{L}_0^\alpha$ (red line and band). Both fits describe the data qualitatively well in the range of lengths studied. We obtain $(c_1,\,c_2)=(0.247\pm0.010, \, 3\pm19)$ and $(A,\,\alpha)=(0.35\pm0.08,\,0.95\pm0.03)$. Similar linear and power-law fits are also obtained 
as $\Delta\tau_{\rm dec}=d_1\tilde{E}_{{\rm str},0}+d_2$ and $\Delta\tau_{\rm dec}=B\tilde{E}_{{\rm str},0}^\beta$, yielding $(d_1,\,d_2)=(0.129\pm0.005,\,-2\pm17)$ and $(B,\,\beta)=(0.14\pm0.04,\,0.99\pm 0.03)$. If the linear fit is assumed, we find a particle emission power 
$\tilde{P}_{\rm part}=\text{d} \tilde{E}_\text{str}/\text{d} \tau=7.8\pm0.3$, independent of the loop length. If we use instead the power-law fit, we obtain $\tilde{P}_{\rm part}\propto \tilde L^\gamma$, with $\gamma = \alpha(1-\beta)/\beta = 0.01 \pm 0.03$, reinforcing the linear hypothesis, on which we focus from here on. We note that if we restrict our analysis to loops with $L_0/r_\text{c}\lesssim 3000$, we obtain instead that $\gamma \sim 0.3$. We believe that this result is driven by finite-volume effects that originate due to the used of loops with lengths much larger than the lattice side.

Fig~\ref{fig:stringdynamics}-(C) shows $\Delta\tau_{\rm dec}/\tilde{L}_\text{av,0}$ as a function of $\tilde L_\text{av,0}$ for  
artificial loops. Fitting data to a power law $\Delta \tau_\text{dec}=A\tilde{L}_{{\rm av},0}^\alpha$, yields $(A\cdot 10^3,\,\alpha)=(21\pm3,\,2.027\pm0.025)$ (type I) and $(A\cdot 10^3,\,\alpha)=(33\pm12,\,1.97\pm0.07)$ (type II). While this fit is valid for all type I loops  studied, independently of $(v_1,v_2)$ and $\delta x$ (indicating small discretization effects),  
it does depend on whether dLC is removed. For example, without removal, we see that for the loops with $(v_1,\,v_2)=(0.3,0.6)$, setting the decay time at the moment of dLC, yields $(A\cdot 10^3,\,\alpha)=(8.3\pm1.6,\,2.16\pm 0.03)$ (see empty circles). Fitting the data
as $\Delta\tau_\text{dec}=B\tilde E_{{\rm av},0}^\beta$, we obtain $(B\cdot 10^3,\,\beta)=(5.4\pm1.0,\,1.988\pm0.026)$ (type I) and $(B\cdot 10^3,\,\beta)=(3.4\pm2.5,\,2.08\pm0.13)$ (type II).  Approximating $\alpha \approx \beta \approx 2$, implies a particle emission power as $\tilde{P}_{\rm part}=C/\tilde L$, with $C \equiv (2\sqrt{A B})^{-1} = 47\pm 5$ (type I) and $C = 47\pm19$ (type II).  Particle production for artificial loops is therefore suppressed the larger the loop. 

The different behavior observed between artificial and network loops is related to their distinctive structure. While a detailed explanation requires a dedicated study of the microscopic mechanisms that drive particle emission, which goes beyond our present analysis, we can still provide some qualitative insights. Artificial loops are composed by a fixed number of kink-like sharp features, joined by straight string segments---see for example \cref{fig:stringsnapshotsartificial}. Network loops, on the other hand, do not possess long straight segments, but are characterized by the presence of a larger number of localized features with nonzero-curvature, as can be observed in \cref{fig:networkevolution}. While the features in network loops are typically not as sharp as kinks, they are still expected to emit particles, as actually seen in~\cite{Blanco-Pillado:2023sap}, while their number grows with the length of the loop. Thus, artificial loops only emit particles from a fixed number of localized features, while network loops emit from features spread from all over their length, consequently decaying at a faster rate. We note this explanation goes in the line of that presented in~\cite{Hindmarsh:2021mnl}. We postpone a proper analysis of these aspects for a future dedicated study of particle and GW emission from the localized features in the loops.

Finally, we highlight that, unlike global loops~\cite{Baeza-Ballesteros:2023say}, the dynamics of 
local loops are very sensitive to 
ultraviolet (UV) effects.  
For example, generating 
the same type I loop configuration with different lattice spacings $\delta\tilde x$, leads to identical $\Delta\tau_{\rm dec}$  for $\delta\tilde x = 0.0417$ and $\delta\tilde x = 0.0625$, but underestimates it by $\sim 5, 10, 15$ and $40\%$ for $\delta\tilde x = 0.125, 0.1875, 0.25, 0.5$, respectively. 
As a compromise, 
we 
use $\delta\tilde{x}=0.1875$ in our study,  guaranteeing $\lesssim 10$\% systematic errors. For network loops, using a coarse-graining procedure as in~\cite{Hindmarsh:2021mnl}, we observe that $\delta\tilde{x}=0.125$ leads to lifetimes only $\sim 5\%$ larger than for $\delta\tilde{x}=0.25$, so we used the latter.

\section{Results on GW emission}\label{sec:GWs}

Tensor perturbations $h_{ij}$ that represent GWs, i.e. transverse and traceless $\partial_i h_{ij} = h_{ii} = 0$, emitted by loops are obtained by solving $\ddot h_{ij} - \vec\nabla^2h_{ij} = 2m_\text{p}^{-2}\Pi_{ij}^{\rm TT}$, where $m_\text{p} \simeq 2.435\cdot 10^{18}$ GeV is the reduced Planck mass, $[ \cdots ]^{\rm TT}$ implies transverse-traceless projection, and $\Pi_{ij} \equiv 2{\rm Re}[(D_i\varphi)(D_j\varphi)^*] - E_iE_j - B_iB_j$. We obtain the (normalized) GW energy density spectrum as~\cite{Caprini:2018mtu}
\begin{eqnarray}\label{eqn:GWpowerspectrum}
\Omega_{\rm GW}(k,t) \equiv  
{1\over \rho_{\rm t}} \frac{\text{d}\rho_{\rm GW}}{\text{d}\log k} =
\frac{k^3m_\text{p}^2}{8\pi^2{\mathcal V}\rho_{\rm t}}\big\langle\dot{{h}}_{ij}(k,t)\dot{h}^*_{ij}(k,t)\big\rangle_{\hat \Omega_k},\nonumber
\end{eqnarray}
where $\rho_{\rm t}$ is the total energy density of the scalar and gauge fields in a lattice of volume $\mathcal{V}$, and $\langle \cdots \rangle_{\hat \Omega_k}$ represents angular averaging in Fourier space. The total GW energy emitted by a loop is obtained via
\begin{equation}
    E_\text{GW}(t) = \rho_{\rm t}{\mathcal V}\int \Omega_{\rm GW}(k,t)\,\text{d}\log k\,,
\end{equation}
so that in program variables we write  
$\tilde\Omega_{\rm GW}(\kappa,\tau) = (v/m_\text{p})^{-2}\Omega_{\rm GW}(k,t)$ and $\tilde E_{\rm GW} = \sqrt{\lambda}\,(v/m_\text{p})^{-2}(E_{\rm GW}/v)$. For simplicity, we focus on the GW emission by network and artificial type I loops only.

 \begin{figure*}[!t]
\includegraphics[width=1\textwidth,height=6.2cm]{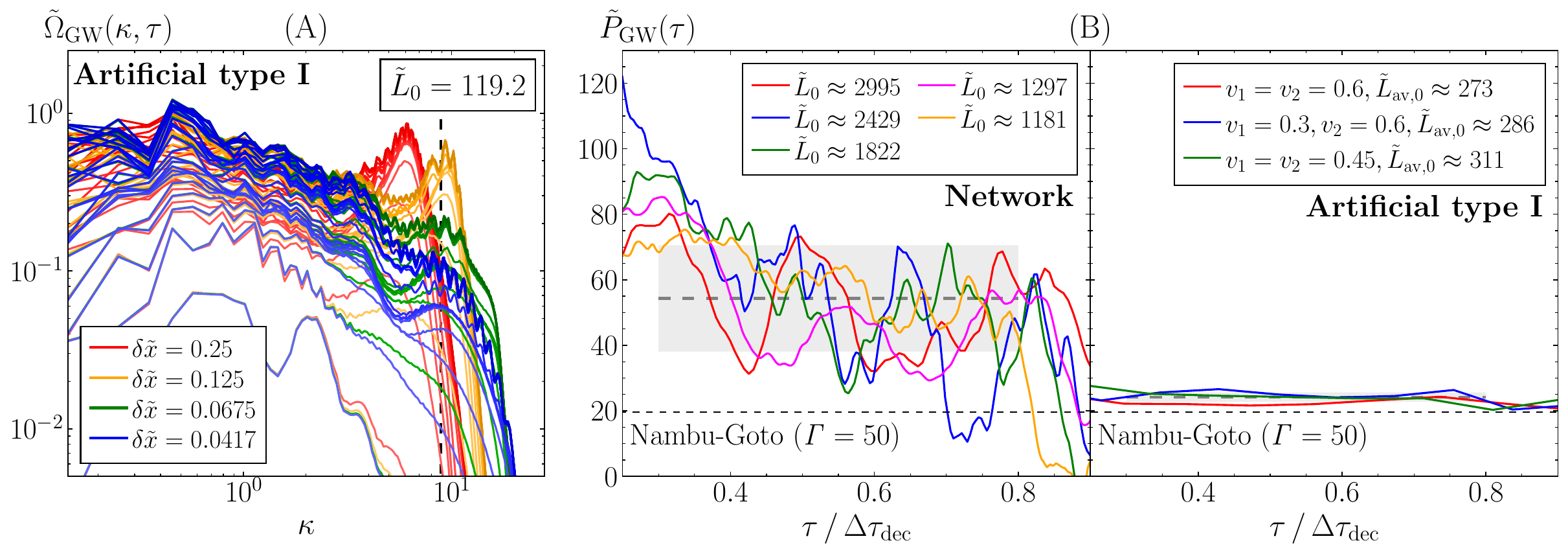}
\caption{Panel (A): Evolution of $\tilde\Omega_{\rm GW}(\kappa,\tau)$  for artificial type I loops with varying UV resolution, $\delta\tilde x$, but fixed lattice size. 
The dashed vertical line indicates the scale of the string width, $\kappa_\text{c} = 2\pi/\tilde r_\text{c}$. Spectra go from early to late times from bottom to top, with lines 
plot every $\Delta\tau = 4$ units. 
Panel (B): Power emission of GWs of network (left) and type I artificial (right) loops. 
}
     \label{fig:UVIRspectra}
 \end{figure*}

Given the UV dependencies observed in 
the dynamics of local loops, we first need to quantify UV effects on their GW emission. In fig.~\ref{fig:UVIRspectra}-(A) we plot the evolution of the GW spectra emitted by a given artificial type I loop configuration, for several values of $\delta\tilde{x}$. We observe the presence of an infrared (IR) peak around $\kappa \sim 0.4-0.5$ for all resolutions. A second peak emerges at UV scales for large $\delta \tilde{x}$ (red and yellow spectra) close to the core radius scale, at $\kappa_\text{c}=2\sqrt{2}\pi$ (dashed vertical line). This UV peak is suppressed for finer lattices (green and blue spectra), signaling it is a lattice artifact. Overall, we observe that simulations with $\delta\tilde{x}\lesssim 0.1875$ agree up to scales $\kappa_\text{cut}\sim 2.5$ for type I loops, which justifies 
to compute $\tilde E_{\rm GW}$ integrating up to such cutoff. 
For network loops we also keep the same cutoff. 

We determine the GW power emitted by network loops as a {\it rolling average}, with 
\begin{eqnarray}\label{eq:averagedGWsPower}
P_{\rm GW}(t) \equiv \frac{\rho_{\rm t}{\mathcal V}}{2T}\int^{t+T}_{t-T}\text{d}t'\int_0^{k_\text{cut}} \dot\Omega_\text{GW}(k,t')\,\text{d}\log k\,,\nonumber 
\end{eqnarray}
which we plot in the left panel of fig.~\ref{fig:UVIRspectra}-(B) 
in terms of program variables, 
$\tilde P_{\rm GW} \equiv (P_{\rm GW}/v^2)\,/\,(v/m_\text{p})^2$,   as a function of the lifetime fraction of the loops $\tau/\Delta\tau_{\rm dec}$, for $\tilde{T} \equiv (\sqrt{\lambda}v) T = 20$. We have checked that the results are insensitive to choosing larger values of $\tilde T$. The amplitude of $\tilde P_{\rm GW}$ is roughly constant, with fluctuations depending on the string evolution.  Averaging on the range $\tau/\Delta\tau_\text{dec}\in[0.3,0.8]$, leads to $\tilde P_{\rm GW} =54\pm16$, which is only $\sim 20\%$ of the value obtained for global network loops~\cite{Baeza-Ballesteros:2023say}, but still $\sim 2-3$ times larger than the NG prediction (shown by the horizontal dashed black line, using $\mu=\pi v^2$ and $\mathit{\Gamma}=50$~\cite{Blanco-Pillado:2017oxo}).

For type I artificial loops, a different approach is required to determine their GW power emission.  This is because, due to their longer lifetimes (often dozens of times the half-box-light-crossing time of the lattice), previously emitted GWs can interfere with newly produced ones, leading e.g.~to spectral oscillations in the late time GW spectra,  see fig.~\ref{fig:UVIRspectra}-(A).  To prevent this volume effect affecting the results, we set GWs to zero when the loop starts oscillating, and evolve the GWs normally until we measure the GW power spectrum after a time $\Delta T_\text{GW}$. We then reset GWs to zero and repeat the procedure until the loop decays. The averaged GW emission power during each sub-interval is computed as
\begin{equation}\label{eq:PowerAverageArtificial}
P_\text{GW}(t)=\frac{\rho_{\text{t}}{\mathcal V}}{\Delta T_\text{GW}}\int_0^{k_\text{cut}}\Omega_\text{GW}(k,t)\,{\rm d}\log k\,.\nonumber
\end{equation}
We use $\Delta\tilde T_{\rm GW} = 160$, which is sufficiently large to capture all relevant frequencies, but small enough to prevent finite-volume effects. We find that using $\Delta \tilde{T}_\text{GW}=80-200$ leads to less than $10$\% discrepancies. We plot $\tilde P_{\rm GW}$ for artificial type I loops in the right panel of fig.~\ref{fig:UVIRspectra}-(B), as a function of the loops' lifetime fraction, $\tau/\Delta\tau_{\rm dec}$. 
We find that the emission power remains mostly constant during the lifetime of the loops. 
Averaging on the range $\tau/\Delta\tau_\text{dec}\in[0.3,0.8]$ leads to $\tilde{P}_\text{GW}=24.2\pm1.4$, slightly above the NG prediction for $\mu=\pi v^2$ and $\mathit{\Gamma} = 50$.

Comparing GW and particle emission rates for artificial loops of type I, as obtained earlier, leads to 
\begin{equation}
{P_{\rm GW}\over P_{\rm part}}=\displaystyle{24.2\pm1.4\over 66 \pm 7}\left({v\over m_\text{p}}\right)^2\left({L\over r_c}\right)^{1+\delta}\,,
\end{equation}
with $\delta = 0.007\pm0.018$.
This implies that GW emission dominates ($P_{\rm GW} > P_{\rm part}$) for {\it large} loops with $L > L_{\rm crit}$, while particle production dominates ($P_{\rm part} > P_{\rm GW}$) for {\it small} loops with $L < L_{\rm crit}$, with $L_{\rm crit}$ a {\it critical length}. Neglecting the small correction $\delta$, so that $P_{\rm GW}/P_{\rm part} \propto L$, 
 we find, from the condition $P_{\rm GW}(L_{\rm crit}) = P_{\rm part}(L_{\rm crit})$,
\begin{eqnarray}
L_\text{crit} \approx 
(2.8 \pm 0.3)\,r_\text{c}\left(\frac{v}{\mpl}\right)^{-2}\,,
\end{eqnarray}
with $r_\text{c}$ the string core radius. Using 
PTA constraints, $v \lesssim 3\cdot 10^{-5}m_\text{p}$, 
leads to a critical length at least as large as $L_{\rm crit}/r_\text{c} \gtrsim 10^{9}$. 
We confirm therefore the prediction in~\cite{Matsunami:2019fss} on the existence of a critical length for type I artificial loops. We obtain however a value $\sim 10$ times larger, which we believe is quite precise, as it is based on a larger loop set and on measuring directly the GW emission from the loops on the lattice (as opposed to using NG predictions). This discrepancy may also be due to the slightly different procedure we use to initialize the infinite strings---see \cite{supplemental}---and the fact  we use a different scheme from~\cite{Matsunami:2019fss} to discretize the equations of motion, which ensures exact conservation of Gauss' law on the lattice---see \cite{Figueroa:2020rrl}.

Similarly, for the network loops we obtain, from the linear fit results, 
\begin{eqnarray}
{P_{\rm GW}\over P_{\rm part}} =
\begin{array}{ll}
\displaystyle{54\pm16\over 7.8 \pm0.3}\left({v\over m_\text{p}}\right)^2 &\hspace*{-5mm}\,,
\end{array}
\end{eqnarray} 
Thus, the ratio $P_{\rm GW}/P_{\rm part}$ is scale-invariant. This implies that, contrary to artificial loops, particle emission dominates at all scales for network loops, as supported by our data with length-to-core width ratios up to $L/r_\text{c} \lesssim 6000$, without indication this may change at longer scales.

\section{Discussion and conclusions}\label{sec:conclusions}

Cosmic string networks are predicted by a variety of field theory and superstring 
early universe scenarios~\cite{Kibble:1980mv,Vilenkin:1984ib,Hindmarsh:1994re,Copeland:2009ga,Copeland:2011dx,Vachaspati:2015cma}, and are expected to create a plethora of observational effects, from CMB anisotropies in the form of power spectra~\cite{Ade:2013xla,Lizarraga:2014xza,Charnock:2016nzm,Lizarraga:2016onn,Lopez-Eiguren:2017dmc} and bispectra~\cite{Figueroa:2010zx,Ringeval:2010ca,Regan:2014vha}, to lensing events~\cite{Vilenkin:1984ea,Bloomfield:2013jka}, cosmic ray production~\cite{Brandenberger:1986vj,Srednicki:1986xg,Bhattacharjee:1991zm,Damour:1996pv,Wichoski:1998kh,Peloso:2002rx,Sabancilar:2009sq,Vachaspati:2009kq,Long:2014mxa,Auclair:2019jip}, and GW emission~\cite{Vilenkin:1981bx,Vachaspati:1984gt,Damour:2000wa,Damour:2001bk,Damour:2004kw,Figueroa:2012kw,Hiramatsu:2013qaa,Blanco-Pillado:2017oxo,Auclair:2019wcv,Gouttenoire:2019kij,Figueroa:2020lvo,Gorghetto:2021fsn,Chang:2021afa,Yamada:2022aax,Yamada:2022imq,Servant:2023mwt,Servant:2023tua}. The relevance of particle production by string loops has been however under debate, since Kibble’s pioneering paper~\cite{Kibble:1976sj}. 

In this Letter we study the emission by local string loops of (scalar and gauge) particles and GWs, considering $i)$ {\it artificial} loops, obtained from the crossing of either straight-boosted (type I) or curved-static (type II) infinite strings, and $ii)$ {\it network} loops, obtained from the lattice simulation of string networks. We find that below a critical length, $L_{\rm crit}$, artificial 
loops decay primarily through particle production, whilst for larger loops GW emission dominates. This agrees 
with~\cite{Matsunami:2019fss}, which showed the existence of a critical length for type I artificial loops. 
We obtain however a value of 
$L_{\rm crit}$ 
which is $\sim 10$ times larger than in~\cite{Matsunami:2019fss}. For network loops, on the contrary, we find that particle emission 
dominates for all loops studied, with length-to-core width 
ratios up to 
$L/r_\text{c} \lesssim 6000$. We observe no indication that this should change for longer loops. GW emission in this case, is suppressed compared to particle production as $P_{\rm GW}/P_{\rm part} \simeq 5\cdot(v/m_p)^2 \ll 1$. If we saturate the CMB bound $v \lesssim v_{\rm CMB} = 10^{-3}m_\text{p}$, it is at most as large as $P_{\rm GW}/P_{\rm part} \lesssim 10^{-6}$, which justifies a posteriori that we neglected the backreaction of the GWs onto the loops.

Extrapolating our results to cosmological scales, 
we expect particle emission to reduce the number density of loops along cosmic history, resulting in a suppression 
of the 
GWB spectrum from a local string network:

$\bullet$ Considering artificial loops, the suppression should only affect the highest frequencies in the spectrum, as these are sourced by early produced loops, which will be small enough to verify $L < L_{\rm crit}$. This effect has ben quantified in~\cite{Auclair:2019jip,Auclair:2021jud} using the value of $L_{\rm crit}$ from~\cite{Matsunami:2019fss}, showing a suppression in the high frequency tail of the GWB above a cutoff, which remains however outside the
range of current and planned GW detectors. As our computation of $L_{\rm crit}$ yields a value $\sim 10$ times larger than in~\cite{Matsunami:2019fss}, we expect to reduce the frequency cut-off scale in the GWB, 
although not enough to make it observable. 

$\bullet$ Considering network loops, we expect particle production to 
essentially suppress the GWB from a network of local strings at all frequencies. 
Assuming the linear fit to the decay of the loops, we obtain that loops will decay a factor $ P_{\rm part}/P_{\rm GW} \sim 10^6 \cdot (v/v_{\rm CMB})^{-2}$ faster than based solely on GW emission. This will reduce substantially the number of loops available at any moment in cosmic history, suppressing  
the GWB amplitude in the whole frequency range. We will present the details of this  
elsewhere. Here we simply anticipate that current constraints on 
GWB signals by PTA observations, will restore the compatibility of the data with energy scales associated to the string formation up to GUT scales, $v \sim 10^{-3}m_\text{p}$, similarly as in the CMB bound. 

As a final remark, we highlight that network loops are precisely the type of loops expected from a scaling local string network generated after an early Universe phase transition, so they represent more realistic configurations than the artificial loops. Moreover, we emphasize that our results for network loops hold even when the core's width represents only $\sim 0.01\%$ of the loop's length, i.e.~in a situation where Nambu-Goto is expected to be very good description of the motion of the loop. Our results show therefore that, just because the motion of the loop's core is well described by Nambu-Goto, it does not mean that particle emission is absent, as is implicitly assumed when describing a Nambu-Goto string.

\section*{Acknowledgments}

 We are grateful to Tanmay Vachaspati for useful conversations. J. B. B. is supported by the Spanish Ministerio de Universidades grant FPU19/04326. J. B. B. and D. G. F. also acknowledge support from projects PID2020-113644GB-I00 and PID2023-148162NB-C21 from the Spanish Ministerio de Ciencia e Innovaci\'on and Agencia Estatal de Investigaci\'on (MICIU/AEI) and the European Regional Development Funds (FEDER), from the European project H2020-MSCA-ITN-2019/860881-HIDDeN and from the staff exchange grant 101086085-ASYMMETRY. D. G. F. is also supported by the grants CIPROM/2022/69, EUR2022-134028, by the Generalitat Valenciana Grant PROMETEO/2021/083, and by Spanish Ministerio de Ciencia e Innovación grant PID2023-148162NB-C22. E. J. C. is supported by STFC Consolidated Grant ST/X000672/1. J .L. acknowledges support from Eusko Jaurlaritza IT1628-22 and by the PID2021-123703NB-C21 grant funded by MCIN/AEI/10.13039/501100011033/ and by ERDF; ``A way of making Europe”. This work has been possible thanks to the computing infrastructure of Tirant and LluisVives clusters at the University of Valencia, FinisTerrae III at CESGA, and MareNostrum 5 at Barcelona Supercomputing Center.

\onecolumngrid
 \appendix

\section{Simulation parameters and results}\label{app:simulationdetails}

This appendix summarizes the simulation parameters and the results for the initial length, energy and decay time of different loops considered in this study. Table~\ref{tab:artificialIC} refers to artificial loops, while table~\ref{tab:networkIC} focuses on network loops.

\begin{table*}[h]
\centering
\renewcommand{\arraystretch}{1.3}
\begin{tabular}{|>{\centering\arraybackslash}p{0.8cm}|>{\centering\arraybackslash}p{0.8cm}|>{\centering\arraybackslash}p{0.8cm}|>{\centering\arraybackslash}p{1.2cm}|>{\centering\arraybackslash}p{1cm}|>{\centering\arraybackslash}p{1cm}|>
{\centering\arraybackslash}p{1cm}|>{\centering\arraybackslash}p{1cm}|}
\hline
\multicolumn{8}{|c|}{Type I} \\
\hline
 $v_1$ & $v_2$ & $\tilde{L}_{\rm B}$ & N & $\delta\tilde{x}$  & $\tilde{L}_\text{av,0}$ & $\tilde{E}_\text{av,0}$ & $\Delta \tau_\text{dec}$   \\ \hline  
0.6 & 0.6 & 126 & 672 & 0.1875 & 272.8 & 620.4 & 1902.4 \\ 
0.6 & 0.6 & 147 & 784 & 0.1875 & 319.9 & 723.0 & 2594.2 \\ 
0.6 & 0.6 & 168 & 896 & 0.1875 & 361.3 & 819.9 & 3322.4 \\ 
0.6 & 0.6 & 189 & 1008 & 0.1875 & 405.9 & 921.7 & 4181.5 \\ 
0.6 & 0.6 & 210 & 1120 & 0.1875 & 451.9 & 1029.3 & 5180.4 \\ 
0.6 & 0.6 & 98 & 784 & 0.125 & 181.9 & 408.4 & 794.0 \\ 
0.6 & 0.6 & 98 & 784 & 0.125 & 239.0 & 540.6 & 1429.9 \\ 
0.6 & 0.6 & 98 & 784 & 0.125 & 296.6 & 678.3 & 2214.0 \\ 
0.3 & 0.6 & 126 & 672 & 0.1875 & 213.8 & 480.7 & 1088.4 \\ 
0.3 & 0.6 & 147 & 784 & 0.1875 & 250.9 & 562.1 & 1603.9 \\ 
0.3 & 0.6 & 168 & 896 & 0.1875 & 286.1 & 643.9 & 2066.4 \\ 
0.3 & 0.6 & 189 & 1008 & 0.1875 & 322.6 & 724.2 & 2721.3 \\ 
0.3 & 0.6 & 210 & 1120 & 0.1875 & 354.0 & 801.1 & 3243.2 \\ 
0.3 & 0.6 & 252 & 1344 & 0.1875 & 408.5 & 936.3 & 4382.9 \\ \hline
\end{tabular}\hspace{1cm}
\begin{tabular}{|>{\centering\arraybackslash}p{0.8cm}|>{\centering\arraybackslash}p{1.2cm}|>{\centering\arraybackslash}p{1cm}|>{\centering\arraybackslash}p{1cm}|>{\centering\arraybackslash}p{1cm}|>{\centering\arraybackslash}p{1cm}|}
\hline
\multicolumn{6}{|c|}{Type II} \\\hline
$\tilde{L}_{\rm B}$ & N & $\delta\tilde{x}$  & $\tilde{L}_\text{av,0}$ & $\tilde{E}_\text{av,0}$ & $\Delta \tau_\text{dec}$   \\ \hline  
126 & 672 & 0.1875 & 213.8 & 480.7 & 1088.4 \\ 
147 & 784 & 0.1875 & 250.9 & 562.1 & 1603.9 \\ 
168 & 896 & 0.1875 & 286.1 & 643.9 & 2066.4 \\ 
189 & 1008 & 0.1875 & 322.6 & 724.2 & 2721.3 \\ 
210 & 1120 & 0.1875 & 354.0 & 801.1 & 3243.2 \\ 
252 & 1344 & 0.1875 & 408.5 & 936.3 & 4382.9 \\  \hline
\end{tabular}\hspace{1.5cm}
\caption{Summary of the simulation parameters (in program units) used to simulate artificial loops of type I (left) and type II (right), together with the results for the oscillation-averaged initial length and energy, and the decay time of the isolated loop obtained after removing the effect from dLC. Type I loops are generated using $\sin\alpha=0.4$.}
\label{tab:artificialIC}
\end{table*}

\begin{table*}[t]
\centering
\renewcommand{\arraystretch}{1.3}
\begin{tabular}{|>{\centering\arraybackslash}p{0.8cm}|>{\centering\arraybackslash}p{1cm}|>{\centering\arraybackslash}p{0.9cm}|>{\centering\arraybackslash}p{0.7cm}|>{\centering\arraybackslash}p{1.1cm}|>{\centering\arraybackslash}p{1.1cm}|>{\centering\arraybackslash}p{0.8cm}|}
\hline
 $\tilde{L}_{\rm B}$ & N & $\delta\tilde{x}$ & $\tilde{\ell}_\text{str}$ & $\tilde{L}_0$ & $\tilde{E}_\text{str,0}$ & $\Delta \tau_\text{dec}$   \\ \hline 
560 & 2240 & 0.25 & 50 & 4144.3 & 7747.7 & 874 \\
700 & 2800 & 0.25 & 70 & 4046.3 & 8288.2 & 959 \\
560 & 2240 & 0.25 & 50 & 3692.7 & 6928.4 & 1004 \\
560 & 2240 & 0.25 & 50 & 3526.0 & 7201.9 & 1054 \\
560 & 2240 & 0.25 & 50 & 3399.7 & 6429.5 & 759 \\
700 & 2800 & 0.25 & 70 & 3320.7 & 6557.8 & 929 \\
560 & 2240 & 0.25 & 50 & 3320.7 & 6343.7 & 944 \\
560 & 2240 & 0.25 & 50 & 3293.3 & 6556.6 & 1114 \\
560 & 2240 & 0.25 & 50 & 3284.3 & 6322.0 & 939 \\
560 & 2240 & 0.25 & 50 & 3152.0 & 6121.0 & 769 \\
560 & 2240 & 0.25 & 60 & 2995.0 & 5737.3 & 749 \\
560 & 2240 & 0.25 & 40 & 2929.7 & 5605.3 & 749 \\
560 & 2240 & 0.25 & 50 & 2845.7 & 5184.0 & 619 \\
256 & 1024 & 0.25 & 15 & 2441.0 & 4511.4 & 474 \\ 
280 & 1120 & 0.25 & 20 & 2429.3 & 4302.9 & 497 \\
240 & 960 & 0.25 & 15 & 2306.0 & 4482.2 & 421 \\ 
288 & 1152 & 0.25 & 20 & 2165.0 & 3991.9 & 454 \\ 
280 & 1120 & 0.25 & 20 & 2160.0 & 4116.1 & 377 \\
256 & 1024 & 0.25 & 15 & 2140.0 & 4238.0 & 414 \\
280 & 1120 & 0.25 & 20 & 2060.0 & 3955.6 & 393 \\
560 & 2240 & 0.25 & 40 & 1828.0 & 3505.6 & 514 \\
280 & 1120 & 0.25 & 20 & 1821.7 & 3457.9 & 443 \\
256 & 1024 & 0.25 & 15 & 1774.0 & 3526.3 & 311 \\ 
384 & 1536 & 0.25 & 40 & 1763.0 & 3463.9 & 430 \\ 
256 & 1024 & 0.25 & 10 & 1741.0 & 3335.8 & 449 \\ 
240 & 960 & 0.25 & 20 & 1738.0 & 3476.7 & 321 \\ 
240 & 960 & 0.25 & 20 & 1727.0 & 3289.9 & 380 \\ 
384 & 1536 & 0.25 & 40 & 1677.7 & 3145.2 & 373 \\ 
224 & 896 & 0.25 & 12 & 1646.0 & 3012.6 & 352 \\ 
256 & 1024 & 0.25 & 15 & 1624.0 & 3055.4 & 385 \\ 
560 & 2240 & 0.25 & 50 & 1593.0 & 3242.7 & 454 \\
280 & 1120 & 0.25 & 20 & 1503.0 & 2698.9 & 397 \\
224 & 896 & 0.25 & 12 & 1476.0 & 2914.1 & 332 \\ 
256 & 1024 & 0.25 & 10 & 1436.3 & 2788.8 & 343 \\ 
256 & 1024 & 0.25 & 15 & 1346.7 & 2588.0 & 440 \\ 
 \hline
\end{tabular}\hspace{1.5cm}
\begin{tabular}{|>{\centering\arraybackslash}p{0.8cm}|>{\centering\arraybackslash}p{1cm}|>{\centering\arraybackslash}p{0.9cm}|>{\centering\arraybackslash}p{0.7cm}|>{\centering\arraybackslash}p{1.1cm}|>{\centering\arraybackslash}p{1.1cm}|>{\centering\arraybackslash}p{0.8cm}|}
\hline
 $\tilde{L}_{\rm B}$ & N & $\delta\tilde{x}$ & $\tilde{\ell}_\text{str}$ & $\tilde{L}_0$ & $\tilde{E}_\text{str,0}$ & $\Delta \tau_\text{dec}$   \\ \hline 
 256 & 1024 & 0.25 & 15 & 1297.3 & 2458.9 & 252 \\ 
256 & 1024 & 0.25 & 15 & 1289.0 & 2282.3 & 304 \\ 
256 & 1024 & 0.25 & 15 & 1224.0 & 2299.2 & 330 \\ 
256 & 1024 & 0.25 & 15 & 1183.0 & 2222.1 & 337 \\ 
256 & 1024 & 0.25 & 10 & 1181.3 & 2327.6 & 359 \\ 
256 & 1024 & 0.25 & 10 & 1181.0 & 2289.3 & 358 \\ 
256 & 1024 & 0.25 & 10 & 1117.7 & 2226.1 & 362 \\ 
256 & 1024 & 0.25 & 15 & 1054.7 & 2068.8 & 262 \\ 
224 & 896 & 0.25 & 12 & 1053.0 & 2075.0 & 217 \\
256 & 1024 & 0.25 & 15 & 1020.7 & 2027.1 & 277 \\ 
128 & 512 & 0.25 & 8 & 1019.0 & 2032.9 & 213 \\ 
256 & 1024 & 0.25 & 15 & 1007.0 & 1993.2 & 369 \\ 
224 & 896 & 0.25 & 12 & 992.0 & 1929.4 & 278 \\ 
256 & 1024 & 0.25 & 15 & 979.3 & 2130.6 & 266 \\ 
256 & 1024 & 0.25 & 15 & 922.0 & 1779.3 & 273 \\ 
224 & 896 & 0.25 & 12 & 920.0 & 1582.0 & 252 \\ 
280 & 1120 & 0.25 & 20 & 874.7 & 1821.4 & 223 \\
128 & 512 & 0.25 & 10 & 848.3 & 1622.7 & 209 \\ 
256 & 1024 & 0.25 & 15 & 817.3 & 1655.0 & 217 \\ 
256 & 1024 & 0.25 & 15 & 793.0 & 1554.3 & 273 \\ 
256 & 1024 & 0.25 & 10 & 776.0 & 1573.9 & 205 \\ 
240 & 960 & 0.25 & 20 & 759.0 & 1748.0 & 196 \\ 
128 & 512 & 0.25 & 12 & 704.7 & 1424.6 & 175 \\ 
256 & 1024 & 0.25 & 15 & 689.0 & 1340.3 & 200 \\ 
128 & 512 & 0.25 & 15 & 684.7 & 1280.5 & 164 \\ 
128 & 512 & 0.25 & 10 & 633.7 & 1305.9 & 139 \\ 
128 & 512 & 0.25 & 15 & 614.7 & 1163.0 & 147 \\ 
128 & 512 & 0.25 & 12 & 612.0 & 1213.8 & 146 \\ 
160 & 1280 & 0.125 & 20 & 607.8 & 1244.7 & 181 \\ 
128 & 512 & 0.25 & 12 & 551.7 & 1118.3 & 142 \\ 
384 & 1536 & 0.25 & 40 & 455.7 & 986.8 & 134 \\ 
128 & 512 & 0.25 & 15 & 432.3 & 843.6 & 127 \\ 
128 & 512 & 0.25 & 8 & 226.3 & 508.7 & 60 \\ 
128 & 512 & 0.25 & 12 & 201.3 & 452.3 & 40 \\ \hline
\end{tabular}
\caption{Summary of the simulation parameters (in program units) used to study the decay of network loops, together with the initial length, energy and decay time of the isolated loop.}
\label{tab:networkIC}
\end{table*}

\twocolumngrid
\bibliography{automatic,manual}

\end{document}